%% file: ms.tex
\documentclass[10pt,sigconf]{acmart}

\copyrightyear{2019} 
\acmYear{2019} 
\setcopyright{acmcopyright}
\acmConference{MobiCom'19}{TBD}{TBD}
\acmPrice{15.00}

\input{preamble}

\renewcommand\footnotetextcopyrightpermission[1]{} 
\def\ca{Carrier A\xspace}
\def\cb{Carrier B\xspace}
\def\kmh{\textit{km/h}\xspace}
\def\tool{\Mod{MPerfDiag}\xspace}

\def\pku{\superscript{\dag}}
\def\umn{\superscript{$\sharp$}} 
\def\hp{\superscript{\S}}  
\def\cas{\superscript{$\flat$}}  
\def\csrs{\superscript{$\natural$}} 

\renewcommand{\baselinestretch}{1}
\definecolor{darkgreen}{rgb}{0, 0.5, 0}
\newcommand{\feng}[1]{{\color{darkgreen}#1}}
\newcommand{\fengc}[1]{{\color{darkgreen}\sout{#1}}}
\newcommand{\BULLET}{\vspace{+.05in} \noindent $\bullet$ \hspace{+.00in}}

\begin{document}

\title{An Active-Passive Measurement Study of TCP Performance over LTE on High-speed Rails}
\titlenote{This work is a pre-print version to appear at MobiCom 2019.}

\author{\Large{{Jing Wang}\pku, {Yufan Zheng}\pku, {Yunzhe Ni}\pku, {Chenren Xu}\pku, {Feng Qian}\umn, {Wangyang Li}\pku, {Wantong Jiang}\pku \\
\vspace{-6pt}
{Yihua Cheng}\pku, {Zhuo Cheng}\pku, {Yuanjie Li}\hp, {Xiufeng Xie}\hp, {Yi Sun}\cas, {Zhongfeng Wang}\csrs}}

\authornote{J.Wang, Y.Zheng and Y. Ni are the co-primary student authors.}

\affiliation{
	\normalsize
	\begin{tabular}{ccc}
	{\pku}Peking University, China & {\umn}University of Minnesota -- Twin Cities, USA & {\hp}Hewlett Packard Labs, USA \\ 
	\end{tabular}
	\\ 
	\vspace{-4mm}{\cas}Institute of Computing Technology, University of Chinese Academy of Sciences \xspace\xspace {\csrs}China Academy of Railway Sciences
}
\input{abstract}



\maketitle

\input{intro}
\input{bg}
\input{meth}
\input{overall}
\input{cubic}
\input{bbr}
\input{improve}
\input{discuss}
\input{related}
\input{concl}

\section*{Acknowledgments}
We are grateful to the reviewers and our shepherd, Dr. Matt Welsh in particular, for their constructive critique and comments, all of which have helped us greatly improve this paper. 
This work is supported in part by 
National Key Research and Development Plan, China (Grant No. 2016YFB1001200), National Natural Science Foundation of China (Grant No. 61802007 and 61672499) and Science and Technology Innovation Project of Foshan City, China (Grant No. 2015IT100095).
\newpage

\balance
\bibliographystyle{unsrt}
\bibliography{meas}

\newpage

\end{document}

%% file: preamble.tex
\usepackage[english]{babel}
\usepackage[utf8]{inputenc}

\newcommand{\AUTHORS}{Authors}
\newcommand{\TITLE}{Title}
\newcommand{\KEYWORDS}{Keywords}
\newcommand{\CONFERENCE}{TBD}

\newcommand{\COMMENTS}{yes}

\newif\ifnice
\nicetrue

\ifnice
\makeatletter
\renewcommand{\subsection}{\@startsection{subsection}{2}{\z@}{-.8ex\@plus -.2ex \@minus -.2ex}{.4ex \@plus .2ex \@minus .2ex}{\normalfont\large\rmfamily\bfseries}}
\renewcommand{\subsubsection}{\@startsection{subsubsection}{2}{\z@}{-.4ex\@plus -.2ex \@minus -.2ex}{.2ex \@plus .2ex \@minus .2ex}{\normalfont\normalsize\slshape}}
\makeatother
\else
\fi

\usepackage{color,balance,appendix,xspace,verbatim,ifthen,gensymb,amsmath,engord}

\usepackage{booktabs,colortbl,diagbox,multirow,tabularx,tablefootnote}

\usepackage{dblfloatfix} 

\usepackage{graphicx,epsfig,epstopdf,subfigure} 
\graphicspath{{./figures/}}
\DeclareGraphicsExtensions{.pdf,.mps,.png,.jpg,.eps,.PNG,.JPG}
\hypersetup{
	plainpages=false,
	filecolor=cyan,
	linkcolor=red,
	citecolor=blue,	
	urlcolor=magenta,
	pdfauthor={\AUTHORS},
	pdftitle={\TITLE},
	pdfsubject={\CONFERENCE},
	pdfkeywords={\KEYWORDS},
	bookmarksopen = {true}	
}

\usepackage{url}
\newcommand{\secref}[1]{\S\ref{#1}}
\newcommand{\figref}[1]{Fig.~\ref{#1}}
\newcommand{\tabref}[1]{Tab.~\ref{#1}}

\newcommand{\figcaption}[1]{\vspace{-0.15in}\caption{#1}\vspace{-0.1in}}
\newcommand{\tabcaption}[1]{\vspace{0.05in}\caption{#1}\vspace{-0.25in}}

\usepackage{tikz}

\newcommand*\circled[1]{\tikz[baseline=(char.base)]{\node[shape=circle,draw,inner sep=0.5pt] (char) {#1};}}

\newcommand{\superscript}[1]{\ensuremath{^{\textrm{#1}}}}

\newcommand{\nosection}[1]{\vspace{3pt}\noindent\textbf{#1.}}

\newcommand{\heading}[1]{\vspace{3pt}\noindent\textup{\textbf{#1}}}
\newcommand{\blpara}{\vspace{3pt}\noindent$\bullet$\hspace{1mm}} 

\def\It{\textit}

\def\eg{\textit{e.g.,}\hspace{1mm}}
\def\ie{\textit{i.e.,}\hspace{1mm}}

\def\etc{\textit{etc.}}

\usepackage{courier} 

\newcommand{\Mod}[1]{\mbox{\textsf{#1}}} 

\usepackage{enumitem}

\newenvironment{Itemize}{
	\begin{list}{$\bullet$} {
			\setlength{\itemsep}{0pt}
			\setlength{\parsep}{2pt}
			\setlength{\topsep}{2pt}
			\setlength{\partopsep}{0pt}
			\setlength{\leftmargin}{1.5em} 
			\setlength{\labelwidth}{1em} 
			\setlength{\labelsep}{0.5em} 
	}}
	{\end{list}}

\ifthenelse{\equal{\COMMENTS}{yes}}{
	\newcommand{\todo}[1]{\textcolor{red}{\textbf{TODO:} #1}}
	\newcommand{\fyi}[1]{\textcolor{blue}{#1}} 
	\newcommand{\fye}[1]{\textcolor{red}{#1}}  
	\newcommand{\remind}[1]{\footnote{\textit{\textcolor{red}{\textbf{Remind:} #1}}}}
	
	\newcommand{\del}[1]{\color{blue} {\sout{#1}}}
	\newcommand{\p}[1]{\vskip 1ex \noindent\colorbox{yellow}{\parbox{\columnwidth}{#1}}\vskip 4pt}
	\newcommand{\note}[1]{\vskip 4ex \noindent\colorbox{yellow}{\parbox{\columnwidth}{#1}}\vskip 6ex} 
	\newcommand{\q}[1]{\vskip 1ex \noindent\colorbox{magenta}{\parbox{\columnwidth}{\textbf{Question:} #1}}\vskip 4pt}
	\newcommand{\qa}[1]{\hl{\textbf{Answer:} #1}}
}{
	\newcommand{\todo}[1]{}
	\newcommand{\fyi}[1]{#1}
	\newcommand{\fye}[1]{}
	\newcommand{\remind}[1]{}

	\newcommand{\del}[1]{}
	\newcommand{\p}[1]{}
	\newcommand{\note}[1]{}

	\newcommand{\q}[1]{}
	\newcommand{\qa}[1]{}	
}

%% file: abstract.tex
\begin{abstract}

High-speed rail (HSR) systems potentially provide a more efficient way of door-to-door transportation than airplane. However, they also pose unprecedented challenges in delivering seamless Internet service for on-board passengers.
In this paper, we conduct a large-scale active-passive measurement study of TCP performance over LTE on HSR. Our measurement targets the HSR routes in China operating at above 300 km/h. We performed extensive data collection through both controlled setting and passive monitoring, obtaining 1732.9 GB data collected over 135719 km of trips.
Leveraging such a unique dataset, we measure important performance metrics such as TCP goodput, latency, loss rate, as well as key characteristics of TCP flows, application breakdown, and users' behaviors.
We further quantitatively study the impact of frequent cellular handover on HSR networking performance, and conduct in-depth examination of the performance of two widely deployed transport-layer protocols: TCP CUBIC and TCP BBR.
Our findings reveal the performance of today's commercial HSR networks ``in the wild'', as well as identify several performance inefficiencies, which motivate us to design a simple yet effective congestion control algorithm based on BBR to further boost the throughput by up to 36.5\%. They together highlight the need to develop dedicated protocol mechanisms that are friendly to extreme mobility.
\vspace{60mm}

\end{abstract}

%% file: intro.tex
\section{Introduction}
Recently, the rapid development of high-speed rails (HSRs) has dramatically changed the way people commute for medium-to-long distance travel. For instance, a train traveling above 300 \kmh potentially provides a more efficient way of door-to-door transportation than airplane. 
To date, 18 countries have developed HSR to connect major cities. In China, the HSR network exceeds 22,000 km in length; in Europe, HSR even travels across international borders \cite{hsr}; in USA, the HSR projects in Texas and California are under construction and expected to finish in the near future \cite{ushsr}.

While such high mobility brings great transportation efficiency, it also poses unprecedented challenges in delivering seamless Internet service for on-board passengers from the trackside broadband radio (\eg LTE) connectivity in a bottom-up fashion -- from error-prone L1/L2 connectivity to misguided TCP. Specifically, as will be shown, the increasing mobility level poses several new challenges:
it degrades the link quality as the Doppler spread increases, increase the BER and reduces the PHY data rate, and hence throttles TCP throughput; from the \It{handover} perspective, handover not only become more frequent, but also are more likely to fail because of the unreliable handover control signal transmission and the tighter timing budget for handover completion due to the train's ultra-high mobility.

Given HSR's short history, its networking is still a relatively new topic.
Existing experimental studies on HSR networking \cite{xiao2014tcp,li2017longitudinal} focus on measuring TCP performance in a controlled setting, and thus lack in-depth cross-layer insights and an understanding of HSR networking performance ``in the wild''.
To bridge such a gap, in this paper, we conduct a cross-layer and large-scale measurement study of TCP performance on HSR. Our measurement targets consist of three popular HSR routes in China operating at above 300 \kmh. More than 150 million passengers travel on these routes annually. The onboard Internet connectivity is provided by multi-carrier LTE, the main-stream mobile access technology for HSR \cite{fxh3}. Through a period of 10 months, we performed extensive data collection through both passive monitoring (at the LTE gateway whose access was provided by China Academy of Railway Sciences) and controlled experiments. To our best knowledge, this is the largest HSR network trace dataset -- 1732.9 GB data collected over 135719 km of trips.
Our measurement consists of four parts as detailed below.

\BULLET (\secref{ssec:tcpperf}) Leveraging such a unique dataset, we begin with measuring important performance metrics for two TCP variants: CUBIC \cite{ha2008cubic} and BBR \cite{bbr}, which are state-of-the-art transport layer solutions that have registered real-world deployment. We found that the extreme mobility of HSR effectively degrades the performance of these protocols, across all metrics. For instance, when the train speed increases from 300 \kmh to 350 \kmh, the average goodput of CUBIC and BBR decreases by 47.5\% and 40.1\%, respectively, due to frequent handover as well as lower PHY rate. Meanwhile, BBR still holds its native property of low(er) RTT, loss rate and bytes-in-flight (when in comparison to CUBIC) in such extreme mobility environment.

\BULLET (\secref{ssec:flowchar}) We then measure key characteristics of TCP flows, application breakdown, and users' behaviors from the on-board passengers' WiFi traffic data. We found that HTTP(S) still dominates the application protocol usage (94.13\%). Among them, more than 95\% of the flows are composed of text, image and application data (rather than audio and video), and they are slow (less than 1 $Mbps$), short (less than 100 KB) and unlikely to finish when its size is above 1 MB.
Another interesting observation is that in HSR networks, the usage patterns are quite distinct (\eg weekends do not necessarily generate more data traffic, and the diurnal patterns are less prominent) as attributed to the unique context of passenger traveling by train. The above findings shed light on improving traffic classification, resource allocation, and cellular infrastructure planning for HSR networking.

\BULLET (\secref{sec:cubic}) Given the importance of handover in high-mobility cellular access \cite{li2017longitudinal}, we next conduct a quantitative in-depth study of the handover impact on HSR networking performance. Specifically, we first develop appropriate taxonomy of handover (\secref{ssec:lte_ho}), and then correlate the lower-layer LTE messages (\eg PHY rate, handover) with TCP's behavior.
We make four key findings.
First, handover occur frequently in HSR -- every 13.7/8.6 seconds on average at 300/350 \kmh.
Second, depending on a handover's type, its performance impact on TCP varies;
despite most handovers being successful, as the mobility level increases, more and longer unsuccessful handovers will appear, leading to more negative impact.
Third, for successful handover, it is typically too short (\ie tens of $ms$) for TCP to react over its normal RTT of hundreds of $ms$.
The fourth finding is what we call the ``near effect'': an unsuccessful handover (typically more than 1 second) can negatively affect data rate over a much longer window (\eg more than 8 seconds) after its actual occurrence.

\BULLET (\secref{sec:bbr}) We next conduct more in-depth exploration of comparative TCP performance. Our key findings include the following. 
First, we find that BBR recovers more smoothly and slower than CUBIC after all type of handover because it has a intrinsically less radical strategy in expanding congestion window and thus more handover-agnostic. Specifically, it performs slightly better after Radio Link Failure (RLF) handover, but much worse after Non-Access Stratum (NAS) recovery than CUBIC. Even though, BBR achieves comparable throughput with CUBIC, but with a much shorter RTT and packet loss rate.
Second, we find that BBR outperforms CUBIC over the connection with higher random loss and thus more carrier-agnostic in our measurement setting. 
Third, BBR is still suboptimal in HSR with high variant RTT due to its excessively conservativeness in $RTprop$ estimation, a critical parameter controlling its sending window.
We managed to achieve 1.36x throughput improvement by simply tuning the strategy of bandwidth probing and adding a stochastic compensation term in $RTprop$ estimation of BBR (\secref{sec:improve}).

\begin{figure*}
	\subfigure[Successful (Type I)]{\includegraphics[width=0.325\linewidth]{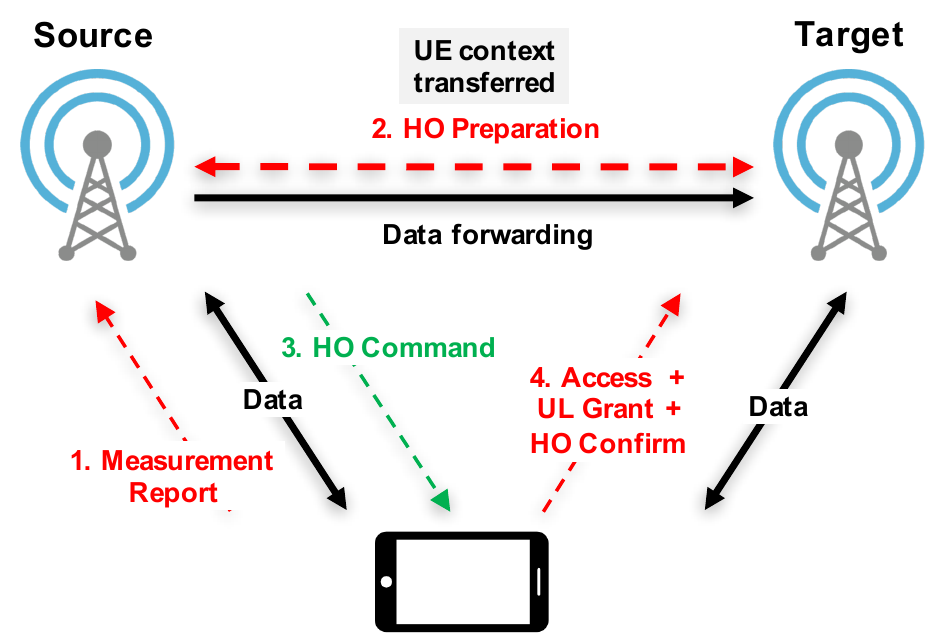}\label{fig:norm_ho}}
	\subfigure[RLF (Type II)]{\includegraphics[width=0.325\linewidth]{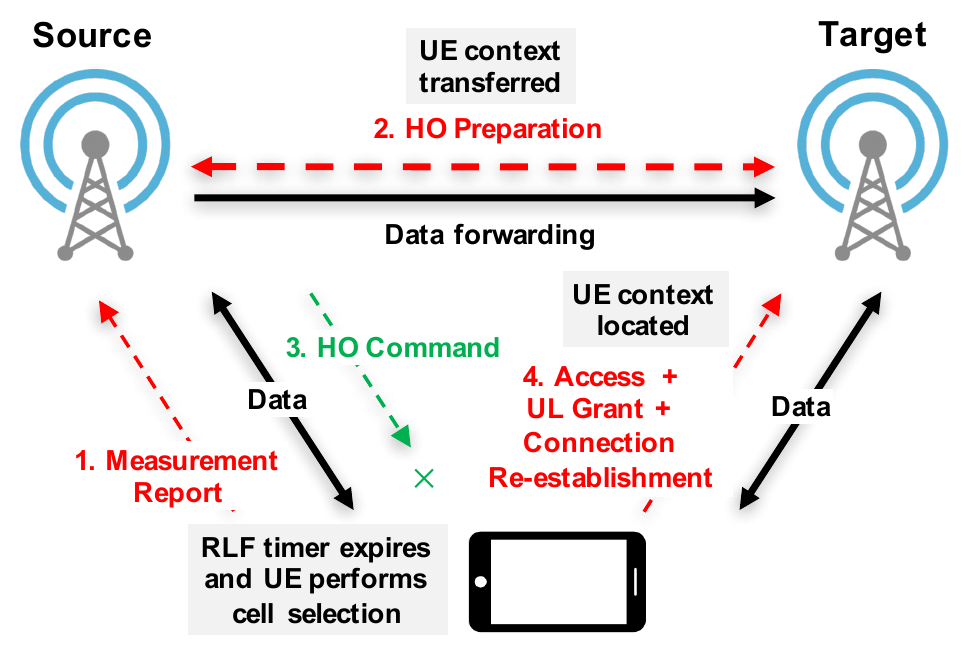}\label{fig:rlf_ho}}
	\subfigure[NAS Recovery (Type III)]{\includegraphics[width=0.325\linewidth]{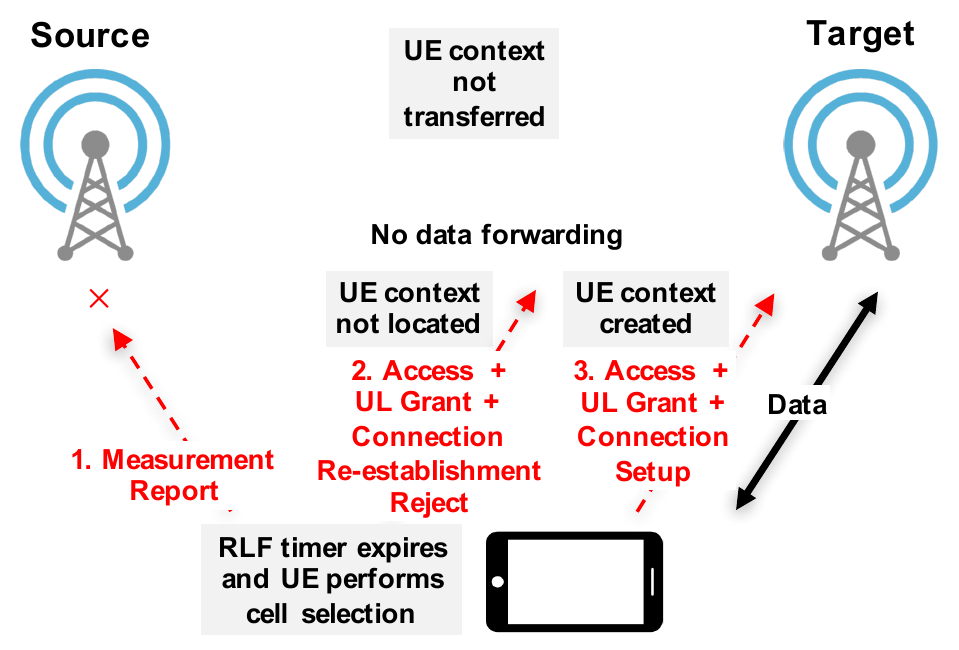}\label{fig:nas_ho}}
	\figcaption{Different LTE Handover Scenarios.}
	\label{fig:ho}
\end{figure*}

\nosection{Contribution} This work represents the first large scale in-depth HSR networking measurement study covering the following aspects: \circled{1} user/traffic behavior in the wild, \circled{2} TCP-LTE interactions, \circled{3} comparative TCP variants behavior, and \circled{4} handover-centric. Here are the key findings:

\begin{Itemize}
	\item HTTP(S) flows dominate on-board Internet traffic, while the QoE (\eg throughput, time to first byte and completion percentage) has huge room for improvement.
	\item Higher train speed not only causes more (unsuccessful) handovers, but also degrades PHY (and TCP) data rate in a non-linear fashion during periods out of handover.
	\item While most handover happen successfully (\ie finish within 100 $ms$) and cause negligible impact on TCP, the unsuccessful handover disrupt TCP in a  complicated manner which depends on the congestion control algorithm.
	\item BBR is more handover and carrier agnostic than CUBIC, but recovers much slower after a long disconnection. We demonstrate there is still great potential for further improvement with a simple end-to-end solution.

\end{Itemize}

As a remark, we believe our study provides key insights for (cross-layer) protocol design dedicated for high mobility data networking in general, and even future standards such as LTE-railway (LTE-R) \cite{he2016high}, a new standard being discussed for the next-generation private HSR communication system for mission-critical services.

%% file: bg.tex
\section{Background}\label{sec:bg}
\subsection{Why LTE is not good enough for HSR}\label{ssec:lteonhsr}
LTE is a 3GPP standard for broadband wireless communication for mobile devices. While it typically provides seamless mobile networking performance for clients on highways or regional trains (of low speed, \ie below 200 \kmh), it runs into severe performance issues when the client mobility is raised to a higher level. According to 3GPP TR 25.913 \cite{25.913}, \It{``Mobility across the cellular network shall be maintained at speeds up 350 \kmh, yet the performance is not guaranteed.''}
There are two major reasons behind it -- poor link quality and frequent handover.

\heading{Link quality} on HSRs becomes poorer than usual mainly because of the large Doppler spread, which is proportional to relative motion between the train and base station. As the mobility level increases, the varying Doppler spread and channel coherence time will incur higher channel estimation errors because of the carrier frequency offset and intercarrier interference \cite{russell1995interchannel,yang2012doppler}. As a result, it not only causes higher decoding errors, but also makes a cell choose lower modulation and coding rate, which together lower the PHY data rate and throttle TCP throughput during the periods even without handover. Another side effect of weak signal is that it reduces the actual on-track LTE coverage, increases the packet loss rate, and hence imposes extra challenges for handover to finish within the overlap zone.

\heading{Handover} on HSRs becomes the main root cause for disrupting a TCP flow -- the increasing mobility level can make it more likely to fail because of the following reasons.
First, as the link quality degrades, the handover control signal might get lost and incur high overhead to recover.
Second, the handover procedure is more likely to fail to finish given the shorter time window within the overlap zone due to high mobility.
Third, the ``tidal effect'' can easily overload the basestation, in both control and data channels.
Upon failure, it needs to spend extra time in discovering and reconnecting to the cell, and thus keeps TCP disconnected.

\subsection{LTE Handover Premier}\label{ssec:lte_ho}
Handover \cite{palat2009lte,dimou2009handover,36.300} is a key function for realizing seamless user experience in mobile networking -- from a source cell to target cell.
In general, a handover can be described as a network-controlled and user equipment (UE)-assisted procedure. According to the 3GPP standard, a handover procedure can be described as follows: UE sends the \Mod{Measurement Report} (\eg signal strength from all the perceived cells) to the source cell.
When the source cell decides to perform a handover, 
it communicates with the target cell for radio resource preparation, informs the UE of the handover action by sending a \Mod{Handover Command} (RRC Connection Reconfiguration) after forwarding UE's buffered downlink (and optionally uplink) user plane data to the target cell for lossless delivery.
Upon receiving this message, UE synchronizes with and gains access to the target cell, and sends a \Mod{Handover Confirm} (RRC Connection Reconfiguration Complete) message to continue the session. In real environment, however, the handover procedure can end up with three different scenarios:

\blpara\It{Successful handover} (\figref{fig:norm_ho}). It happens when all the controlling signals are received, and the buffered data are losslessly forwarded to target cell and delivered to UE for minimizing the flow disruption.

\blpara\It{Radio Link Failure (RLF) handover} (\figref{fig:rlf_ho}).
It happens when radio conditions not good enough for the UE to be able to decode the \Mod{Handover Command} from the source cell. When the UE detects radio link problems, it starts the RLF timer. 
Upon expiration of the RLF timer, the UE searches for a suitable target cell and re-establishes its connection with it (performing the RRC Connection Reestablishment procedure) if it (the target cell) happens to have been prepared by the source cell. RLF handover incurs additional delay, but no data buffered in cell is lost.

\blpara\It{Non-Access Stratum (NAS) recovery} (\figref{fig:nas_ho}): It happens when the target cell is not prepared for handover. In such case, the UE attempts to establish a new connection -- the UE context needs to be created, and all the buffered data is lost and needs upper-layer (TCP) retransmission.

\subsection{TCP Primer}\label{ssec:tcp_review}
We briefly introduce the necessary background on how CUBIC and BBR deals with the network dynamics.

\heading{CUBIC} modifies the linear window growth function of existing TCP standards to be a cubic function. When a loss event happens, CUBIC registers the current congestion window ($cwnd$) as $\textrm{W}_{max}$ and performs a multiplicative decrease of $cwnd$ by a scaling factor. 
The cubic function is set to have its plateau at $\textrm{W}_{max}$ and its increasing is based on elapsed time instead of reception of ACK -- thus the window growth is independent of RTT and flows grow their $cwnd$ at the same rate. After CUBIC enters into congestion avoidance from fast recovery, it starts to increase the window using the concave profile of the cubic function until $cwnd$ becomes $\textrm{W}_{max}$. After that, the cubic function turns into a convex profile to ensure that the window increases very slowly at the beginning and gradually increases its growth rate to probe aggressively for additional capacity. This style of window adjustment (\ie concave and then convex) makes the $cwnd$ remain almost constant around $\textrm{W}_{max}$, improves network utilization and scalability of TCP over fast and long distance (\ie large bandwidth-delay product) networks, and meanwhile treats other TCP connections fairly. However, the fact that it treats packet loss over a lossy wireless link as the signal of network congestion will still throttle its $cwnd$ by mistake thus leads to low bandwidth utilization.

\heading{BBR} employs two parameters, namely $\mathit{RTprop}$ (\ie round trip propagation time  (estimated by taking the minimum RTT over the last 10 seconds) and $\mathit{BtlBw}$ (\ie bottleneck bandwidth  estimated by taking the maximum throughput over the last 10 $\mathit{RTprop}$), to model the end-to-end capacity and determine its congestion control window. Specifically, BBR first uses the slow-start akin to CUBIC's only when the flow is initially launched and then soon reach its bandwidth probing phase after the throughput converges. In this phase, it takes a period-8 cycling \It{$\mathit{pacing\_gain}$} sequence ($1, 1, 1, 1, 1, 1, 5/4, 3/4, \dots$) in turn as a multiplier to $\mathit{BtlBw}$ to determine its sending rate for $\mathit{RTprop}$ (not RTT) time -- while $\mathit{pacing\_gain} = 1$ at most of the time, a $\mathit{pacing\_gain} > 1$ means BBR is in the phase of exploring more bandwidth, after which a $\mathit{pacing\_gain} < 1$ is necessary to guarantee that the queue at the bottleneck will be drained in case there is no more bandwidth to utilize. The take away message is that, BBR is robust to random packet loss, but will take long time (in comparison to CUBIC) to recover after a long  disconnection, \eg after $10 \cdot \mathit{RTprop}$ or longer.

%% file: meth.tex
\section{Measurement Methodology}\label{sec:meth}

\begin{table*}
	\centering\scriptsize
	\begin{minipage}{\linewidth}
		\begin{tabularx}{\linewidth}{|l|X|l|l|l|} \hline
			Factor to examine & Data collection & Experiments & Distance & Data size \\ \hline\hline
			TCP-LTE interaction in high mobility & TCP trace from both ends and LTE trace from client at different speed & Active measurements & 51367 $km$ & 357.9 GB \\ \hline
			TCP variants behavior comparison & TCP trace from servers using different TCP congestion control algorithms & Active measurements & 51367 $km$ & 323.0 GB \\ \hline
			User profile of Internet usage & TCP trace from train-mounted LTE gateway over a few months & Passive measurements & 84352 $km$ & 1376 GB \\ \hline
		\end{tabularx}
	\end{minipage}
	\tabcaption{Dataset description.}
	\label{tab:exp}	
\end{table*}

\subsection{Factors to Examine}
\begin{figure}
	\centering
	\includegraphics[width=\linewidth]{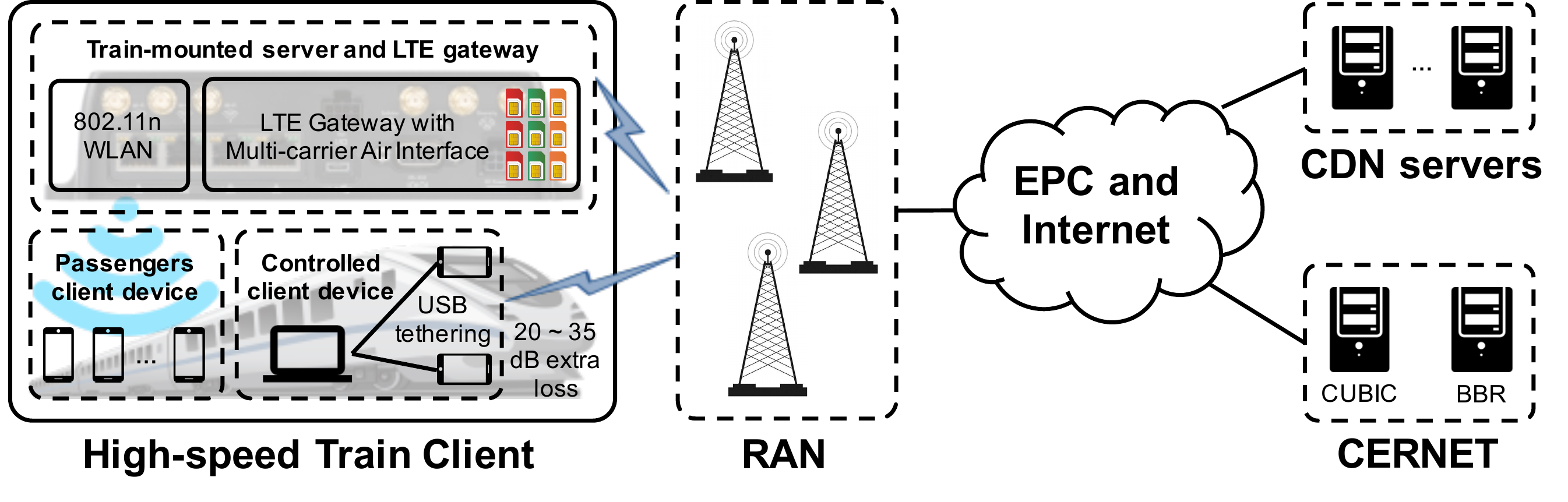}
	\figcaption{Our experimental testbed collects data from both a dedicated laptop-phone suite and a LTE gateway in active and passive manners respectively.}
	\label{fig:exp}
\end{figure}

In order to demystify the performance issues and optimization opportunities in HSR networking, we use real-world data complemented with on-board controlled experiments to gain insights in the following dimensions: 
\circled{1} What is unique about the interaction between TCP and LTE in high mobility environment?
\circled{2} How do different TCP congestion control algorithms behave?
\circled{3} How do the on-board passengers use Internet?
While TCP performance, cross-layer interaction, and user behaviors under stationary or low/moderate mobility have been extensively studied in the literature, there are much fewer studies of them under extreme mobility.
We summarize the high-level experimental design in \tabref{tab:exp} and experimental setup for data collection in \figref{fig:exp}, which will be further explained in the rest of this section.

\subsection{Active Measurements in Controlled Setting}\label{ssec:ctrl_exp}
We begin with describing controlled experiments conducted on HSR. They allow us to capture and analyze a wide range of information in a cross-layer manner.

\subsubsection{Experimental setup}
\nosection{Server} We deploy two powerful co-located servers (Intel NUC6i7KYK with i7-6770HQ, 32 GB DDR4 and Samsung 950 pro 512 GB) in CERNET \cite{li2011china}, the nationwide education and research computer network in China.

\nosection{Client} We tether two Android phones (Xiaomi 5s) to one laptop (Dell XPS 13-9360) via USB 3.0.
This tethered setup allows us to programmatically run two experiments simultaneously on the two phones, which appear as network interfaces on the laptop and function as link-layer devices.
We modified the tethering code in Android OS to provide such multihoming support.
The phones are equipped with SIM cards of two mobile carriers in China, denoted as \ca and \cb.
Note that in China there are three mainstream carriers and we cover two of them; the third carrier uses the same technology (FDD) and exhibits similar performance as \cb based on our pilot on-board tests.

\subsubsection{Experimental design}
Our high-level experimental methodology is to perform bulk data download over TCP. We next detail several important design aspects in terms of what and how to measure.

\nosection{Flow size} We measure two types of TCP downlink flows, including long flows (\ie fixed duration of 150 seconds) and fixed size of 64 KB (corresponding to typical web page). Given the smaller protocol overhead, we believe the TCP behavior in long flows presents a wide variety of Internet contents, ranging from several MB (\ie web contents such as image and JavaScript) to tens of MB such as HD video chunks. These are typical workload of HSR networking.

\nosection{TCP-LTE interaction in high mobility} We run \Mod{tshark} on both client and server to collect packet-level TCP traces. We also instrument the client phones using \Mod{MobileInsight} \cite{li2016mobileinsight}\footnote{Android-based in-device software tool that collects runtime network information and exposes protocol messages on both control plane and (below IP) data plane from the 3G/4G chipset from operational cellular networks.} to collect lower layer information including
PHY rate and handover events. 

\nosection{TCP variants comparison} The two co-located servers run Ubuntu 17.04 with kernel 4.10.17 with CUBIC and BBR, arguably the most widely deployed congestion control algorithms, respectively. We compare their performance and their incurred cross-layer interactions.

\begin{figure}
	\centering
	\includegraphics[width=\linewidth]{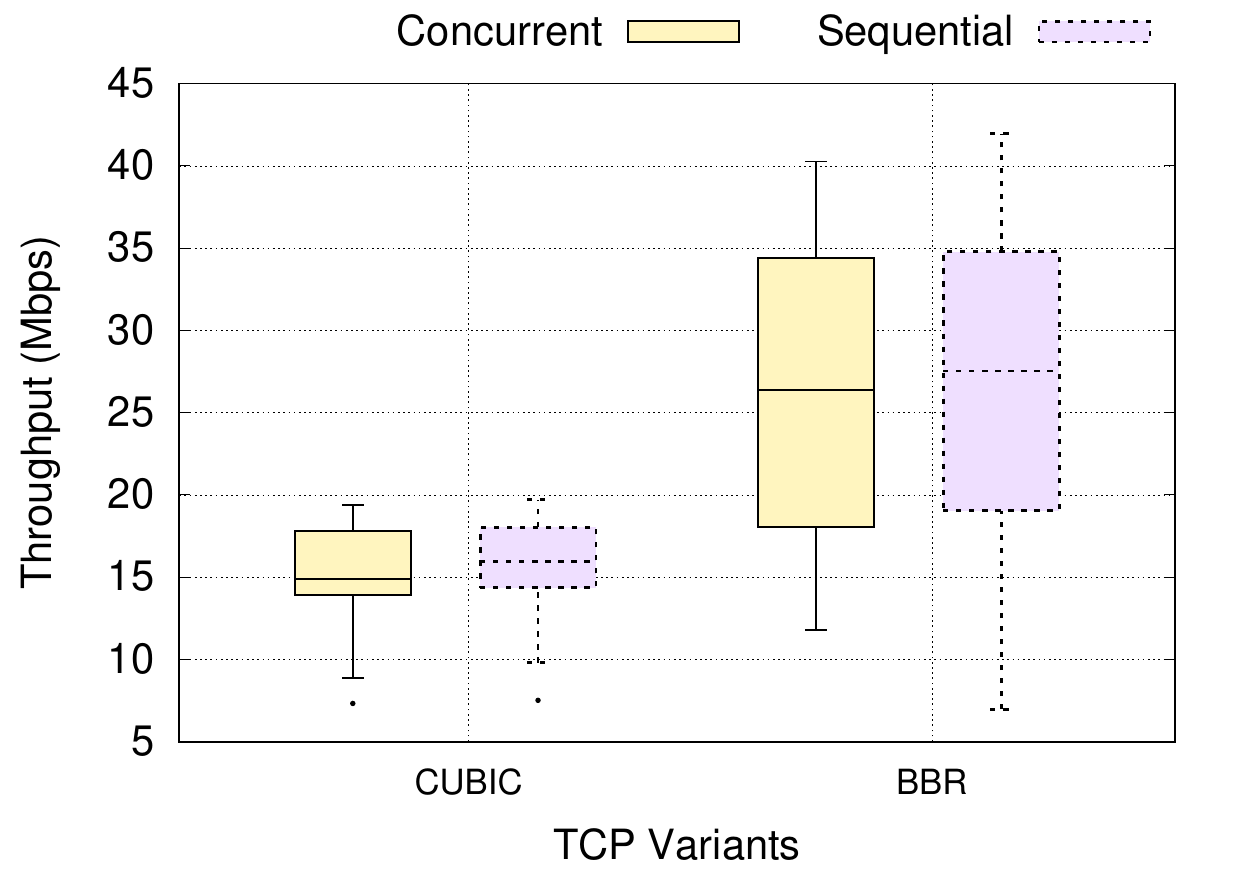}
	\figcaption{Fairness Validation for Concurrent Tests.}
	\label{fig:fair}
\end{figure}

When comparing CUBIC and BBR, we can either execute them sequentially (back-to-back) or concurrently (side-by-side). For experiments under zero or low mobility, one can typically do back-to-back runs. In HSR, however, the channel condition and link quality may change dramatically over a just few seconds' window, so back-to-back runs may not ensure apple-to-apple comparisons.
We therefore run both flows concurrently. 
However, a concern raised here is that whether this side-by-side setting will cause these two flows to interfere with each together over the LTE network. 
To study this, we perform the concurrent and sequential experiments in an interleaved manner for 100 times and run each experiment for 1-minute long.  
We then measure their throughput in \figref{fig:fair}.
As shown, CUBIC and BBR yield qualitatively similar performance when running with and without another concurrent flow. This is likely attributed to the base stations' proportional fair scheduling \cite{kwan2009proportional} as well as their resource capacity capable of serving hundreds of UEs. 
In short, we believe that on a fully provisioned HSR route, the performance impact due to the inter-device contention is dwarfed by the impact caused by the extreme mobility.

\subsubsection{Data Collection and Processing}\label{sssec:activedat}
We carried out experiments on the Beijing-Shanghai (300/350 \kmh) HSR route as it represents the state-of-art HSR networking environments in terms of train speed and track-side cellular infrastructure. We collected 357.9 GB data by traveling 51367 $km$ on the trains. 
Since TCP-LTE performance may vary along the route because of the terrain diversity \cite{luan2013fading} and LTE cell density, we collected the data over the whole route without temporal or spatial sampling. We note that one straightforward way to eliminate the impact of this factor is to log the GPS reading to perform location-aware analysis. However, in our experiments the phone failed to report GPS data at most of the time due to magnetic-shielding from the sealed carriages.

After obtaining this unique dataset, we performed numerous types of data processing such as extracting TCP flows and LTE events, calculating various performance metrics, and aligning TCP traces with LTE events. We next describe how we extract two important types of LTE data.

\blpara\It{PHY Rate} is number of Transport Block (TB) size (\ie the number of bytes that can be carried over a subframe) per second. Specifically, the TB size is jointly determined by the number of resource blocks (RB)\footnote{RB is time-frequency resource that occupies 12 subcarriers (12 $\times$ 15 $kHz$) and one slot (0.5 $ms$). RBs are allocated by the eNB scheduler and this allocation information is sent to the UEs for informing the radio resource and PHY layer rate assignment.} and the modulation/coding scheme (MCS).

\blpara\It{Handover} can happen in three different ways (\secref{ssec:lte_ho}) on HSR-LTE. In the rest of the paper, we refer to the three handover scenarios as Type I, II, and III as shown in \figref{fig:ho}, and denote the start and end time of a handover as $HO_{\textrm{start}}$ and $HO_{\textrm{end}}$, respectively. $HO_{\textrm{end}}$ can be simply determined by the time when Reconfiguration Connection Reconfiguration Complete message is sent (and logged). However, it is more complicated to determine $HO_{\textrm{start}}$, especially for unsuccessful handover -- in practice we don't have the access to the information of RLF timer, which is triggered by radio link failure and leads to cell selection (and handover). Given the streaming nature of our controlled experiment, we hence set the timestamp of the last LTE downlink packet was perceived before the nearest $HO_{\textrm{end}}$ as $HO_{\textrm{start}}$.

We release the dataset used for this study in \cite{hsrnetdat}.

\subsection{Passive Measurements in the Wild}\label{sssec:crowds_exp}
We complement controlled experiments with passive measurements that collect TCP flows from on-board passengers, in order to study the passengers' network usage and their flow characteristics ``in the wild''. We are unaware of any prior passive measurement of HSR networking.

Since 9/2017, China Railway Corporation launched the new ``Fuxing Hao'' trains for the Beijing-Shanghai HSR route. They bring two notable features: cruising at the speed up to 350 \kmh \cite{fxh1} that is faster than any other HSR route in China, and providing free WiFi service via an on-board LTE gateway \cite{fxh2}. Hence, the LTE gateway becomes an ideal point for our passive data collection in the wild.

The LTE gateway is deployed in a on-train server room by China Academy of Railway Sciences (CARS). It runs OpenWRT 3.9, and is equipped with 9 SIM cards of three major Chinese mobile carriers for data relay between LTE RAN and on-board WiFi users. It also deploys a 2 $\times$ 2 MIMO antenna mounted on the top of the carriage. Each new TCP flow is assigned to a SIM card in a round-robin manner.

We obtained permission from CARS to run instrumentation software on the LTE gateway for passive data collection. 
We deployed \Mod{tshark} to collect packet-level TCP traces (headers only) from the LAN port of the gateway.  
Note that we cannot run \Mod{MobileInsight} or other PC-based cellular performance monitoring tool such as QXDM \cite{qxdm} because of the OS incompatibility. But we were able to
distinguish different passengers' TCP flows from their assigned WLAN IP addresses appearing in the PCAP traces.
Overall, we collected 1376 GB data covering 84352 $km$. 

\nosection{Ethical Considerations} We take measures as much as we can to protect users' privacy. 
First, all passengers who participated in our study were presented with informed consent statement before connecting to the on-board WiFi service managed by the HSR operator.
Second, when collecting the traffic trace, only TCP/IP headers were examined, and the clients' IPs are private addresses from which no personal identifiable information could be inferred.
Third, this study has been approved by the Institutional Review Board at the primary authors' institution.

%% file: overall.tex
\section{Basic Performance Statistics}\label{sec:mr}
\subsection{Performance of TCP Variants}\label{ssec:tcpperf}
We first utilize the controlled measurement data to study key network-level performance metrics including goodput, bytes-in-flight (BiF), round trip time, packet loss rate, and out-of-order delay. In particular, we investigate how TCP congestion control algorithm (CCA) affects the above metrics.
\begin{figure*}
	\begin{minipage}[b]{0.655\linewidth}
		\centering	
		\subfigure[300 \kmh]{\includegraphics[width=0.49\linewidth]{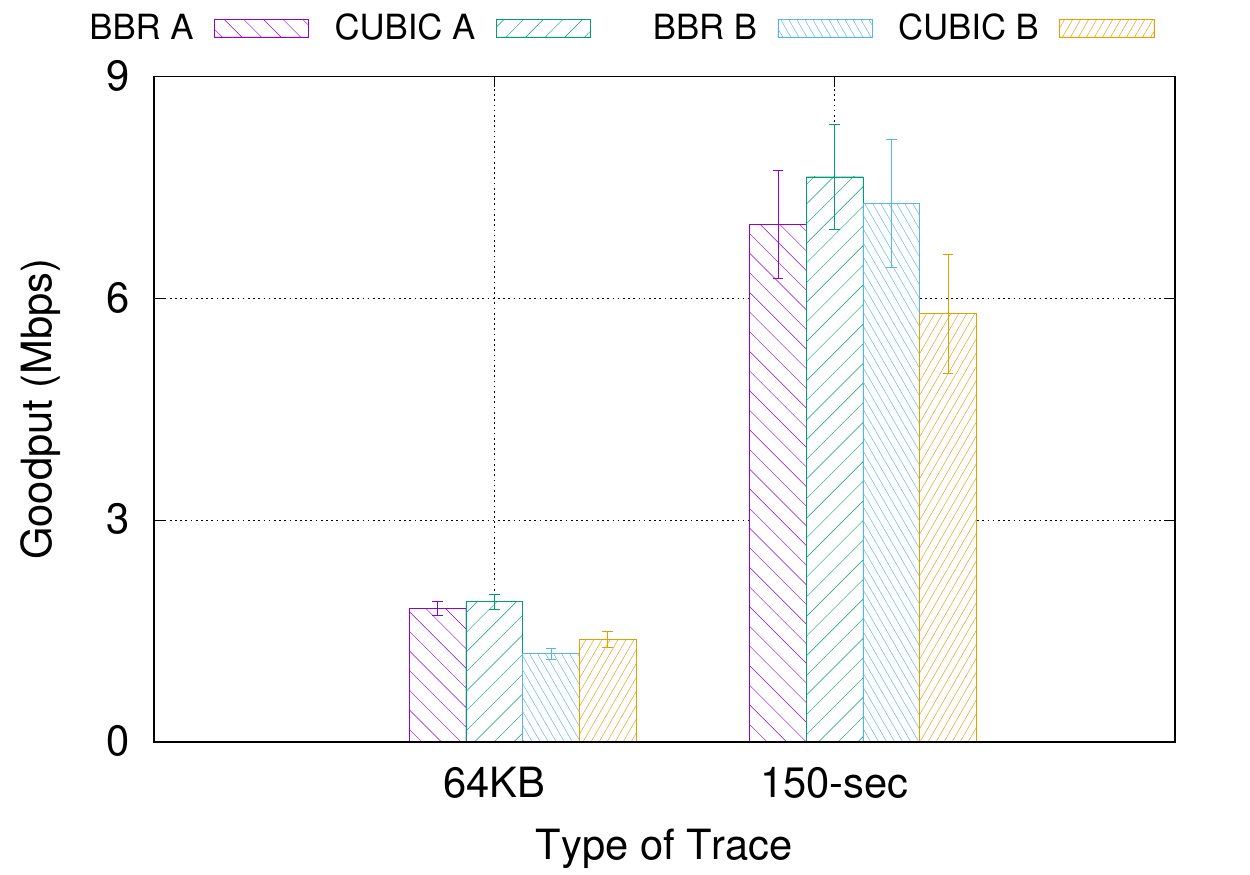}\label{fig:goodput_300}}
		\subfigure[350 \kmh]{\includegraphics[width=0.49\linewidth]{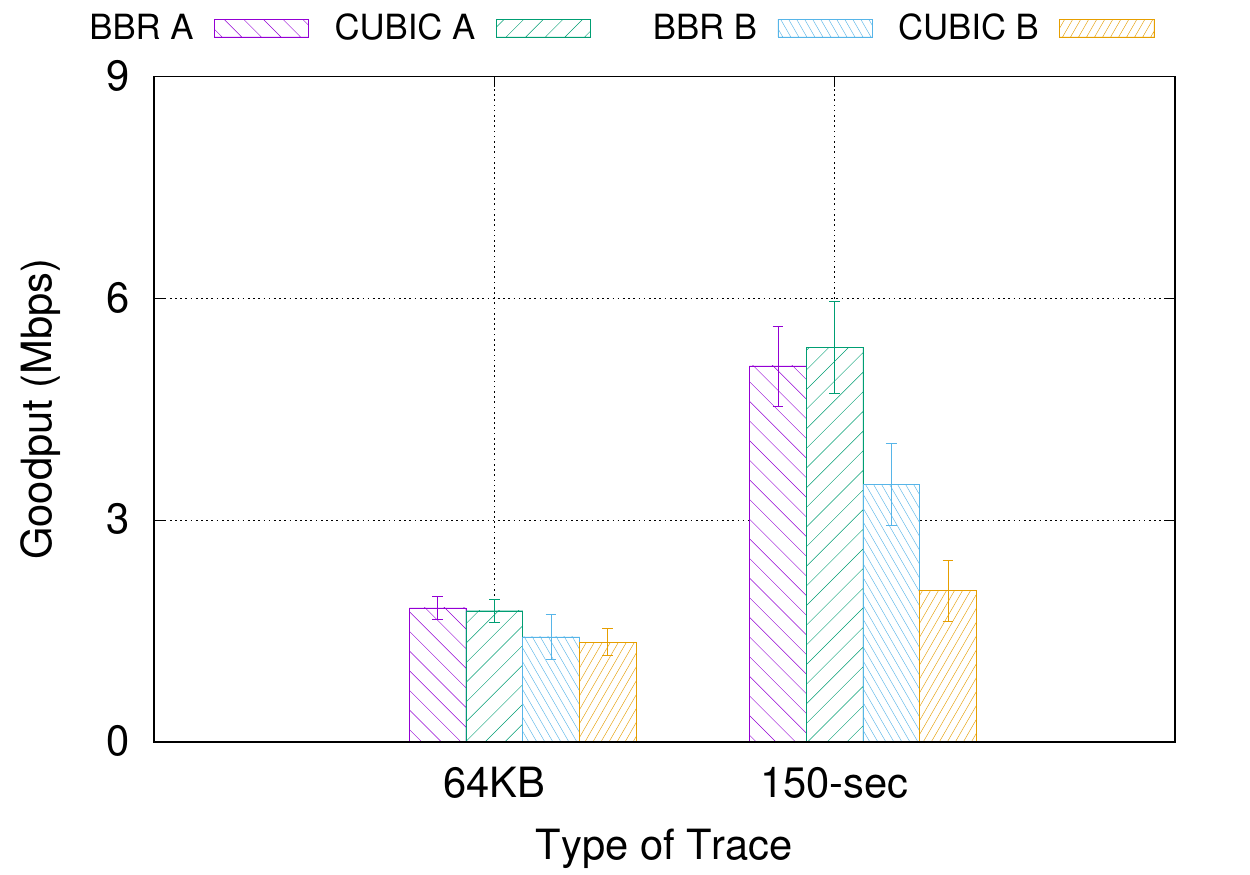}\label{fig:goodput_350}}
		\figcaption{Goodput.}
		\label{fig:goodput}
	\end{minipage}
	\begin{minipage}[b]{0.34\linewidth}
		\centering
		\includegraphics[width=\linewidth]{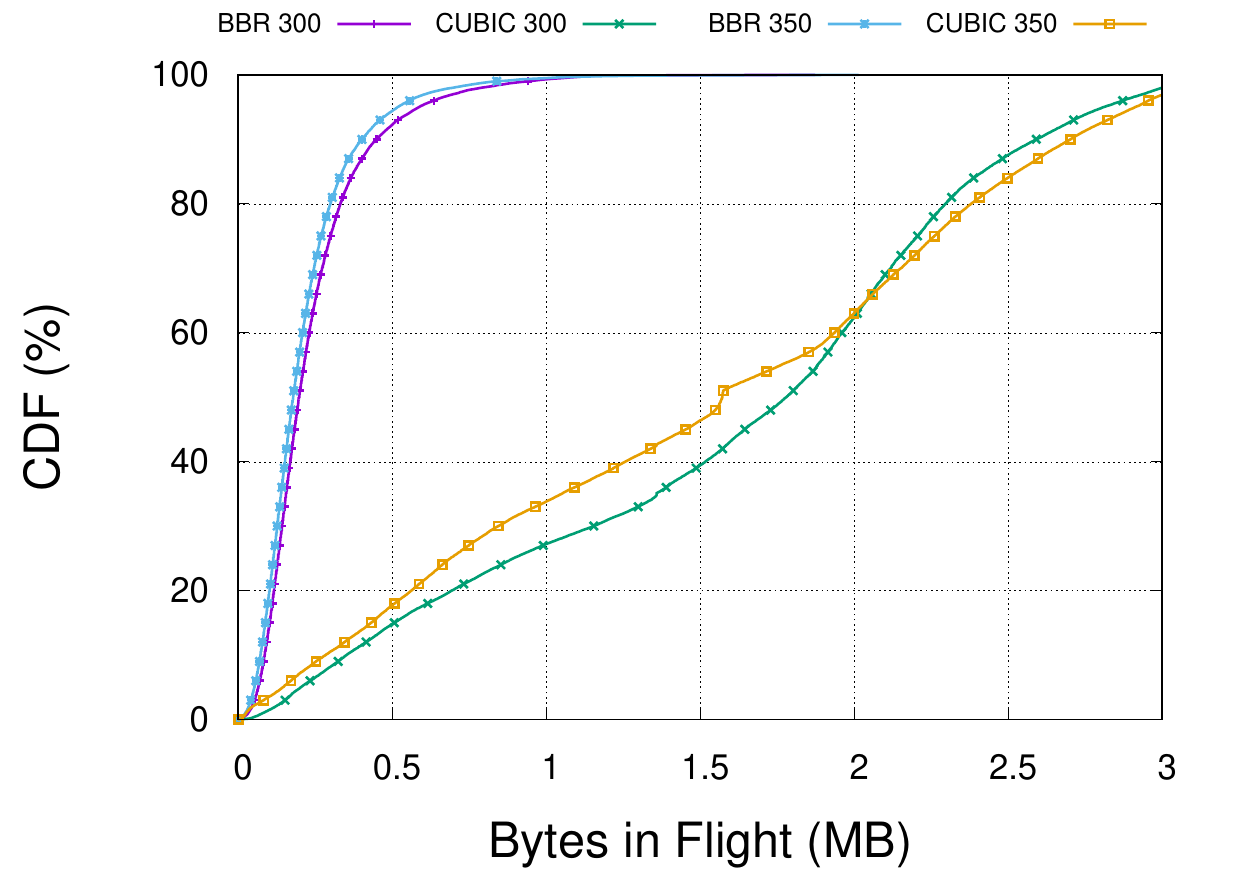}		
		\figcaption{BiF (\ca).}
		\label{fig:bif}
	\end{minipage}	
\end{figure*}
\begin{figure*}
	\begin{minipage}[b]{0.33\linewidth}
		\centering
		\includegraphics[width=\linewidth]{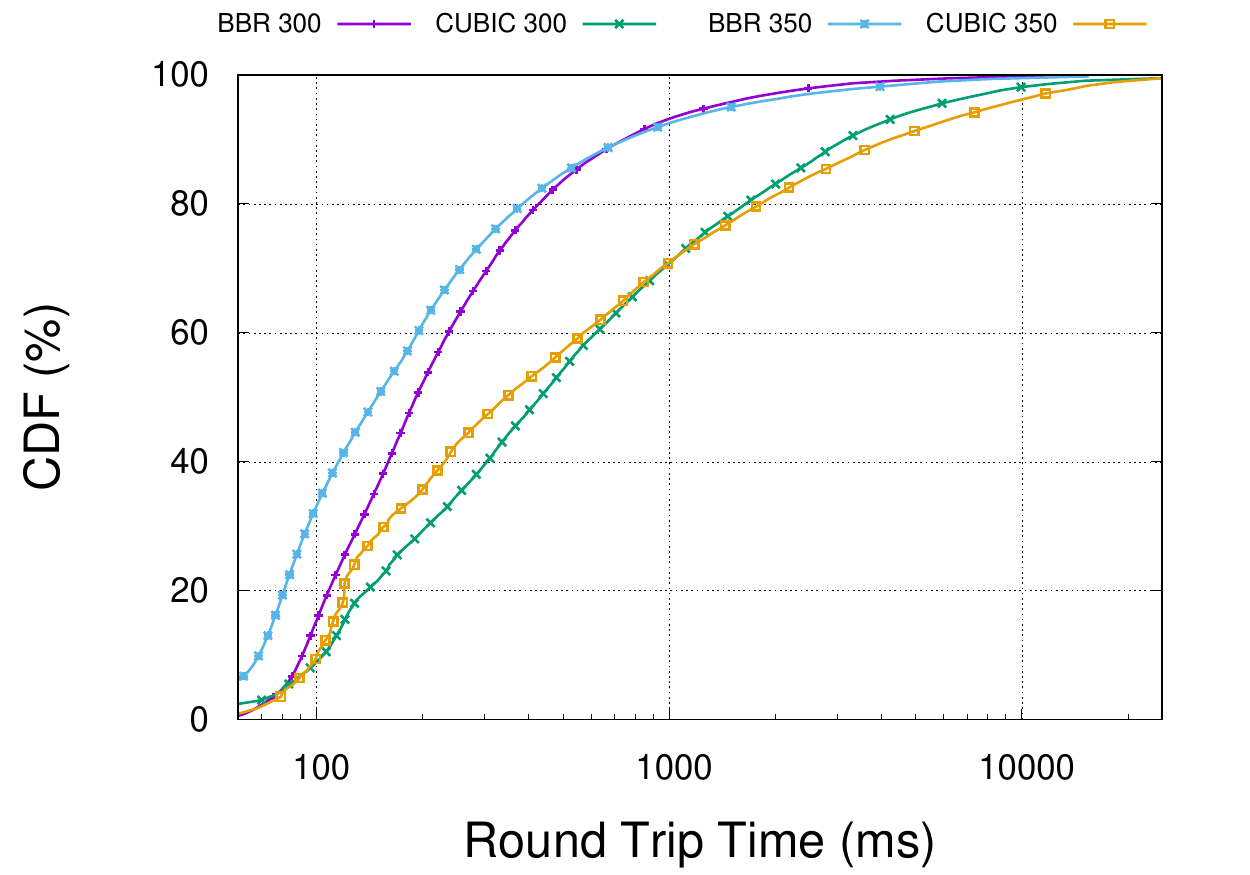}
		\figcaption{RTT (\ca).}
		\label{fig:rtt}		
	\end{minipage}
	\begin{minipage}[b]{0.33\linewidth}
		\centering
		\includegraphics[width=\linewidth]{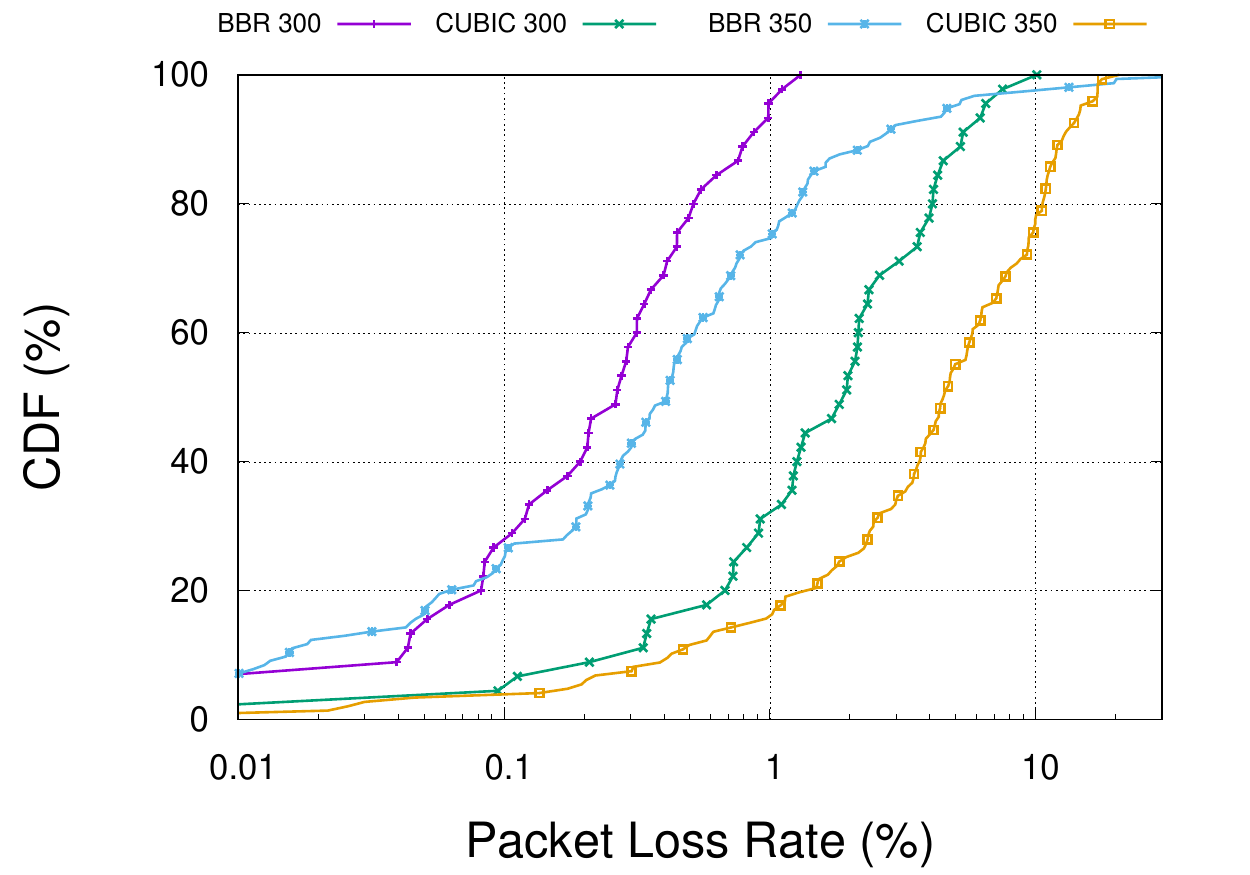}
		\figcaption{PLR (\ca).}
		\label{fig:lr}
	\end{minipage}
	\begin{minipage}[b]{0.33\linewidth}
		\centering
		\includegraphics[width=\linewidth]{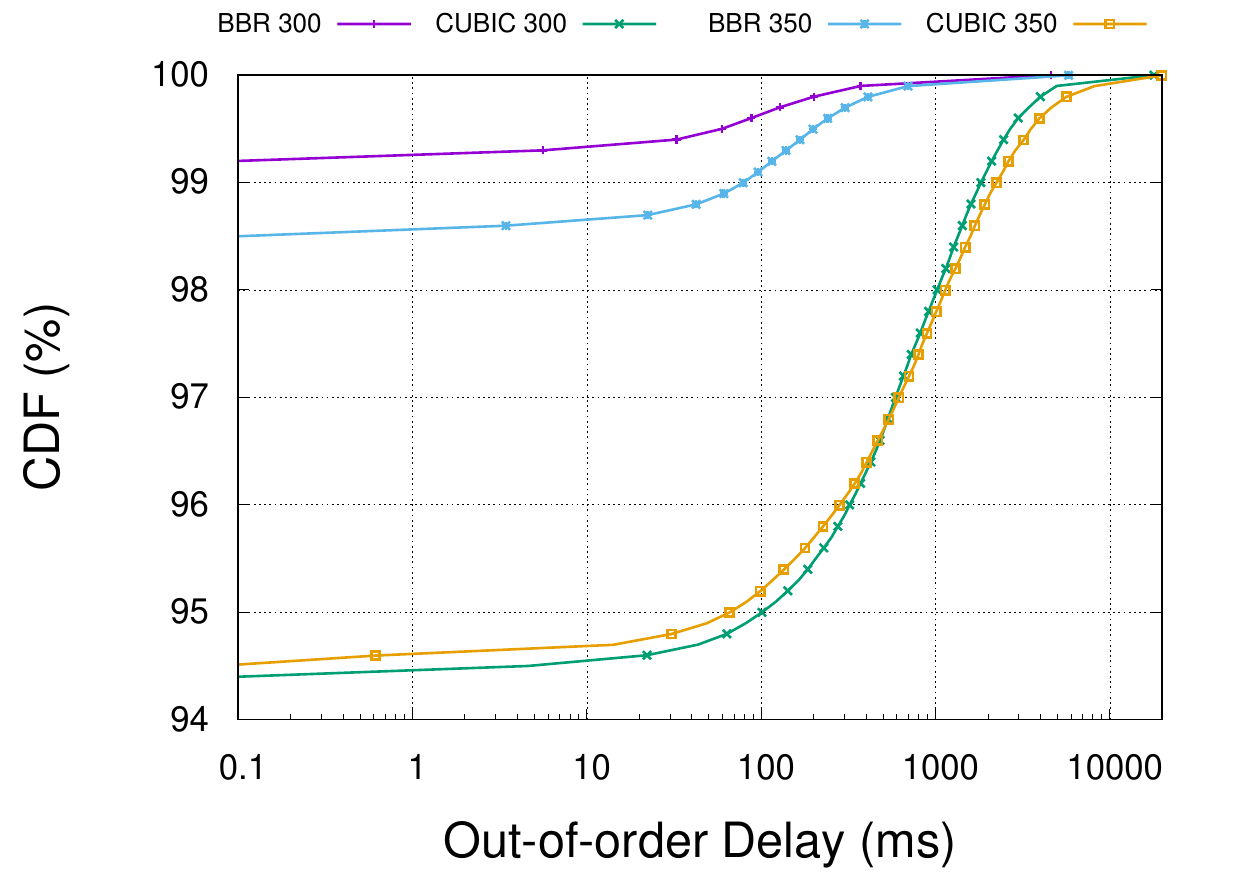}
		\figcaption{OOD (\ca).}
		\label{fig:oood}
	\end{minipage}
\end{figure*}

\nosection{Goodput} 
\figref{fig:goodput} plots the goodput of downloading different files under two speeds (300 \kmh and 350 \kmh) for \ca and \cb. We consider two workloads: a short flow (64 KB) and a long-lived bulk download flow lasting for 150 seconds.
As shown, neither the CCA nor the carrier appears to significantly affect the performance of the short flow, which mostly finishes within the slow start stage during which the available bandwidth is under-utilized.
For the long flow (150 $sec$), we make two key observations.
First, as the mobility level increases from 300 \kmh to 350 \kmh, the goodput of CUBIC and BBR both decrease by 47.5\% and 40.1\%, respectively. This is attributed to the lower PHY rate (caused by the imperfect radio receiver design in high mobility) to be discussed in \secref{sec:cubic}.
Second, when compared to CUBIC, BBR yields marginally lower goodput over \ca, as CUBIC is known to expand its congestion windows (and bytes-in-flight) aggressively. Over \cb, however, BBR yields higher goodput (25.79\% higher at 300 \kmh and 70.19\% higher at 350 \kmh) compared to CUBIC.
This is because \cb has higher random loss rate (to be discussed in \secref{ssec:bbr_perf}) which is infrastructure-dependent. Such random losses force CUBIC to (more) frequently back off while bringing much smaller impact on BBR, which does not rely on random packet losses for modeling the network capacity.

For the sake of space, we will focus on \ca for the rest of the metrics not only because both carriers exhibit similar pattern in terms of comparative performance across CCA and mobility level, but also \ca is the most popular local carrier. We summarize the critical statistics for both carriers in \tabref{tab:perf_stats} by the end of this section.

\begin{table*}[b]
	\centering\tiny
	\begin{minipage}{0.495\linewidth}
		\begin{tabularx}{\linewidth}{|l|l|l|l|X|} \hline 
			Metrics & Mean 	& Median  & 95\% percentile & 99\% percentile \\ \hline\hline
			Goodput (Mbps) & 5.12/5.40 	& 4.97/4.78 	& 11.85/13.06		& 14.60/16.29	\\ \hline	
			BiF (MB) 	 &0.22/1.53  		& 0.18/1.57	& 0.52/2.91	& 0.84/3.14			\\ \hline
			RTT (ms) 	& 457.42/1707.2 		& 148.63/345.02 					& 1491.7/8272.1			& 6022.9/18760.7	\\ \hline
			PLR (\%) 	& 1.38/5.92 		& 0.41/ 4.53		 			& 4.64/14.79			& 20.18/17.77	\\ \hline
			OOD (ms) 	& 4.44/77.75		& 0/0 					& 0/66.62			& 79.79/2244.2	\\ \hline						
		\end{tabularx}
		\vfill\vspace{2mm}
		\centering\footnotesize (a) \ca
	\end{minipage}\hfill
	\begin{minipage}{0.495\linewidth}
		\begin{tabularx}{\linewidth}{|l|l|l|l|X|} \hline 
			Metrics 		& Mean 			& Median  			& 95\% percentile 	& 99\% percentile \\ \hline\hline
			Goodput (Mbps) 	& 3.54/2.06 	& 2.65/1.02 		& 11.20/7.58		& 12.29/10.24 \\ \hline	
			BiF (MB) 		& 0.28/1.08 	& 0.19/1.06 		& 0.82/2.62			& 1.66/3.01	\\ \hline
			RTT (ms) 		& 269.91/2193.3	& 141.42/1067.86	& 798.41/7454.55	& 2095.70/18934.9 \\ \hline
			PLR (\%) 		& 2.13/3.99 	& 0.64/2.25 		& 8.93/11.47 		& 16.04/24.57	\\ \hline
			OOD (ms) 		& 42.47/95.62	& 0/0				& 2.56/170.08		& 740.82/2565.08	\\ \hline			
		\end{tabularx}
		\vfill\vspace{2mm}
		\centering\footnotesize (b) \cb	
	\end{minipage}
	\tabcaption{Critical statistics of BBR/CUBIC performance metrics over different carriers at the speed of 350 \kmh.}
	\label{tab:perf_stats}
\end{table*}

\nosection{Bytes-in-flight (BiF)}
As shown in \figref{fig:bif}, BBR yields an almost a order lower BiF than CUBIC (\eg 0.20 versus 1.78 MB at 300 \kmh, and 0.18 MB versus 1.57 MB at 350 \kmh for median value).
This cross-validates the RTT difference between BBR and CUBIC shown in \figref{fig:rtt}, as a large BiF incurs high queuing delay that inflates the RTT \cite{jiang2012tackling}.
As the mobility level increases, the BiF oftentimes decreases due to reduced throughput. In fact, we found that the higher mobility only causes marginal impact on both CUBIC and BBR. However, we observe that for CUBIC, the BiF can sometimes increase to 3 MB at 350 \kmh. This is explained by the higher likelihood of an uplink ACK packet being delayed or lost, causing a ``spuriously inflated'' BiF. 

\nosection{Round-trip-time (RTT)}
As shown in \figref{fig:rtt}, BBR has more than twice lower RTTs than CUBIC (\eg 191.53 ms versus 431.35 ms at 300 \kmh, and 148.63 ms versus 345.02 ms at 350 \kmh for median value)
due to their different CCA design rationales: BBR intends to suppress the RTT to overcome the bufferbloat problem \cite{gettys2011bufferbloat}.
The increase of mobility level affects the RTT in two aspects. On one hand, more frequent handover and higher packet loss rate (\figref{fig:lr}) lengthen the RTT, especially contribute longer tails; on the other hand, when traveling faster, the CCA dictates the server to send data slower, which oftentimes leads to reduced the in-network buffer occupancy level (as well as queuing delay) and henceforth the RTT. Note that low RTT (\eg less than 200 $ms$) is critical to meet the QoE requirement for popular network applications such as teleconferencing and gaming for on-board passengers.  

\begin{figure*}
	\centering
	\subfigure[Content-type
	]{\includegraphics[width=0.33\linewidth]{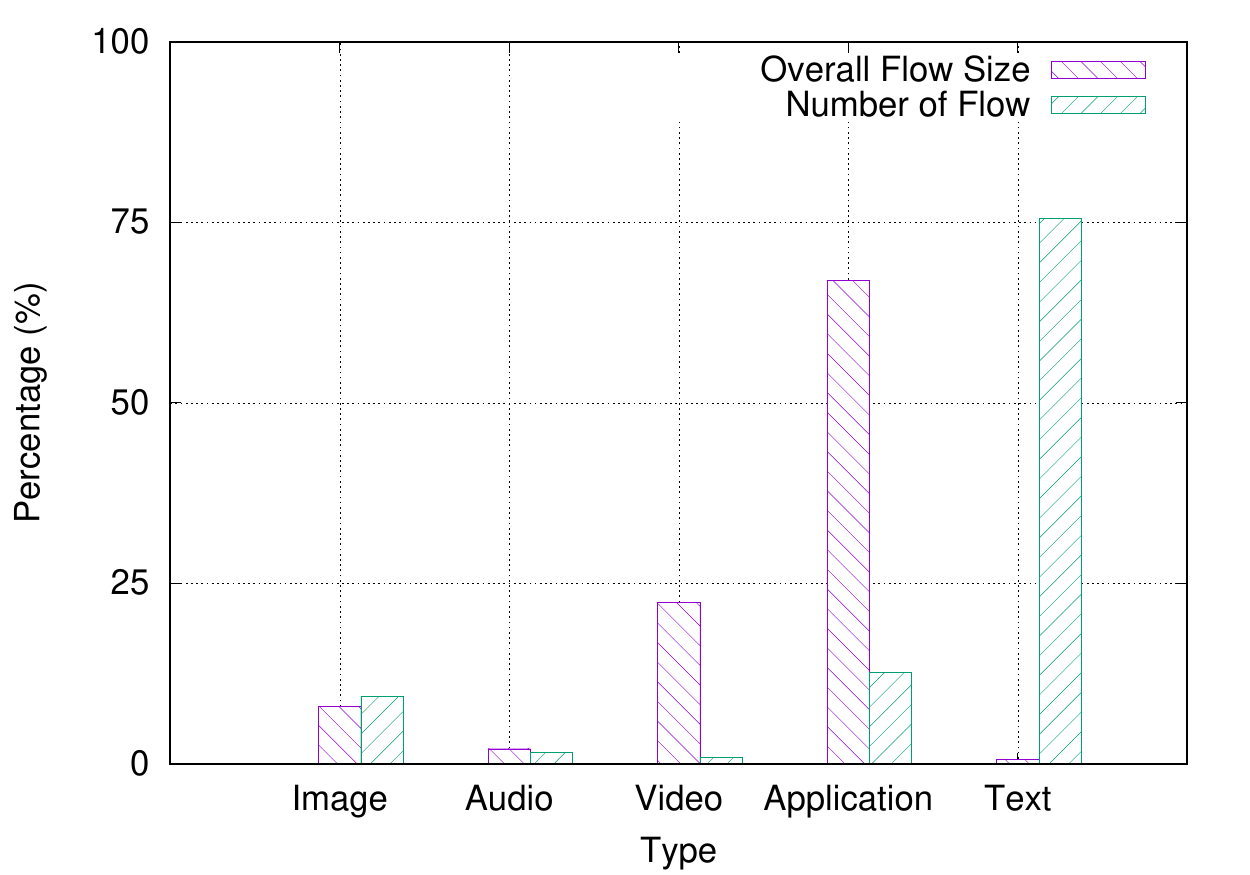}\label{fig:content_type}}
	\subfigure[Object Size]{\includegraphics[width=0.33\linewidth]{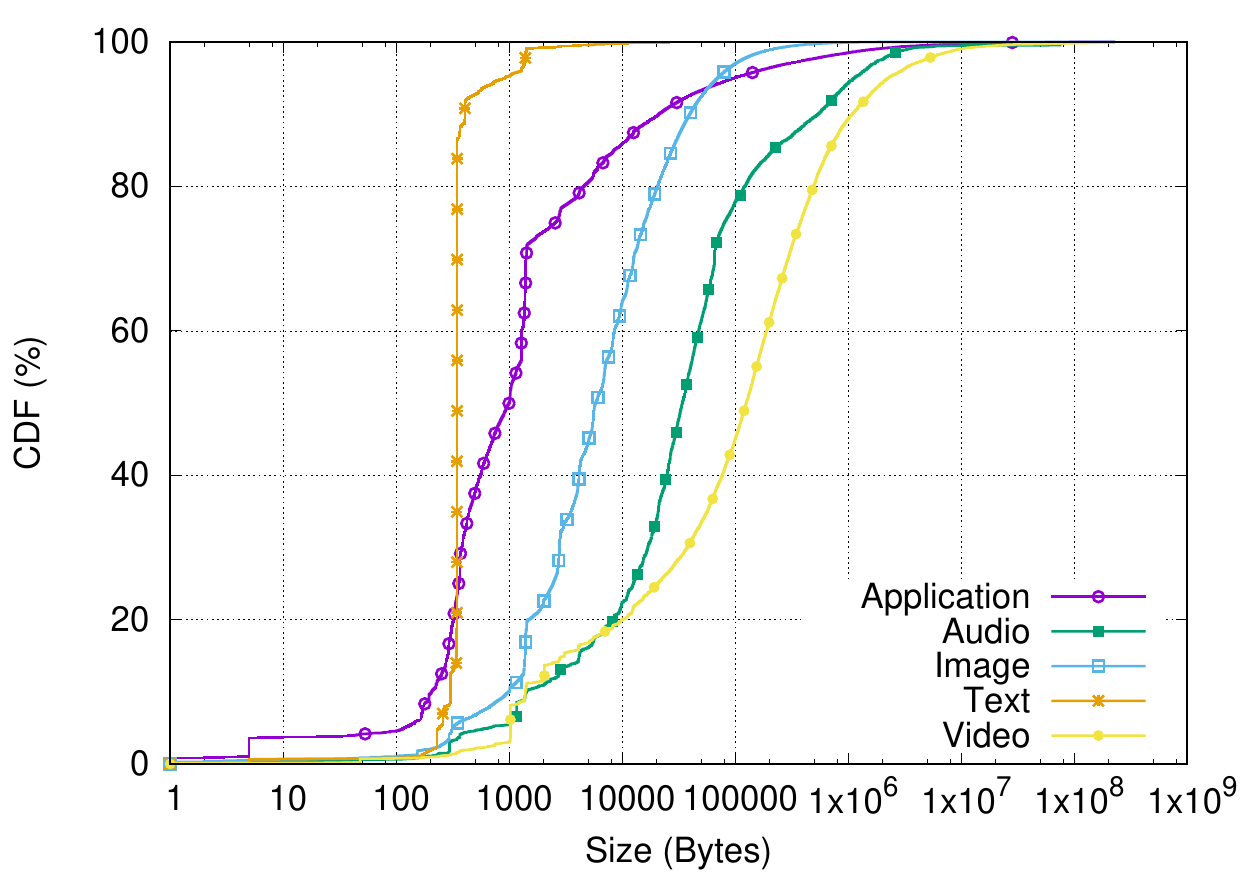}\label{fig:http_flowsize}}
	\subfigure[Object Data Rate for Objects $\ge$ 16 KB
	]{\includegraphics[width=0.33\linewidth]{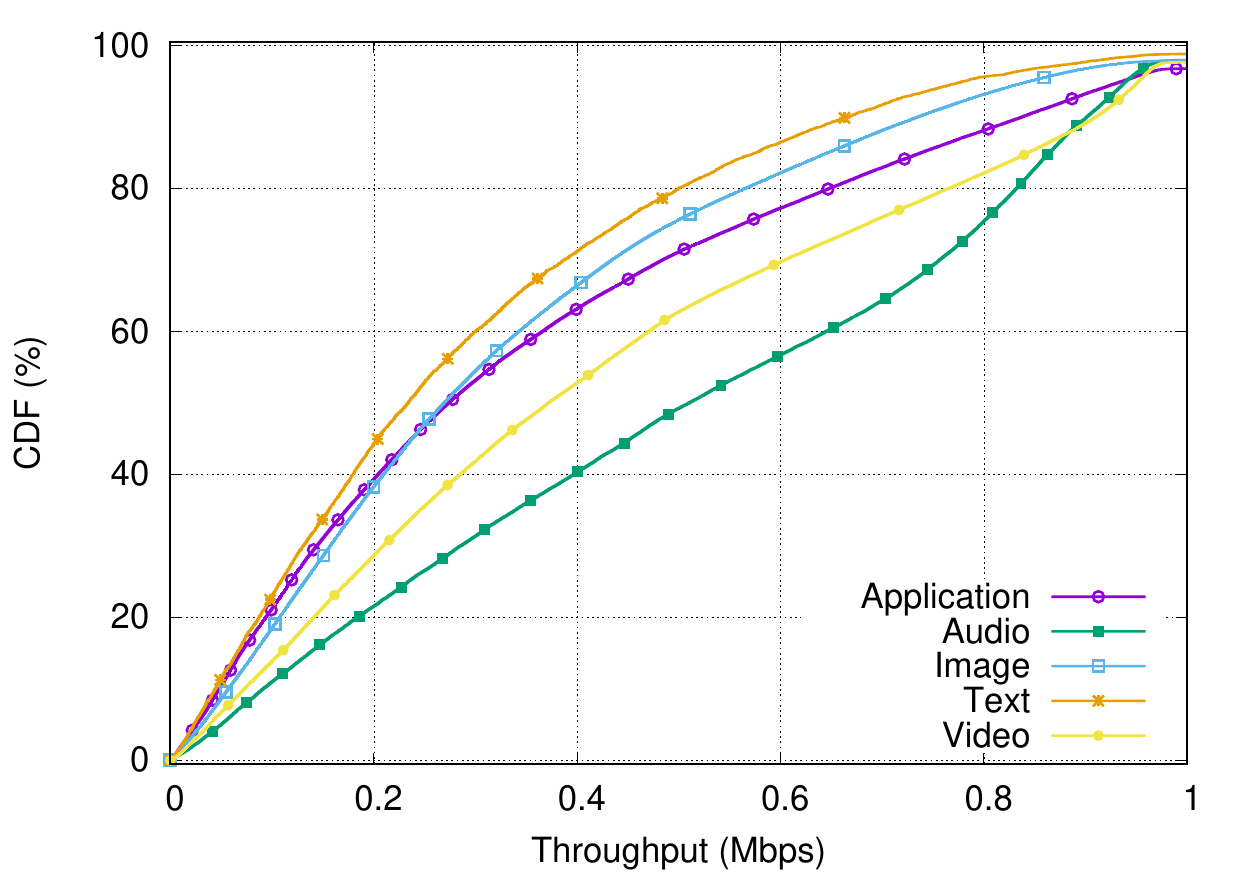}\label{fig:http_thp}}
	\figcaption{Web Content Profile.}
\end{figure*}

\nosection{Packet Loss Rate (PLR)}
As shown in \figref{fig:lr}, BBR has about an order lower PLR than CUBIC (\eg 0.27\% versus 1.95\% at 300 \kmh, and 0.41\% versus 4.53\% at 350 \kmh for median value).
This is because BBR is designed to keep RTT or queuing delay low to avoid tail-drop in the buffer inside the network. As the mobility level increases, PLR increases because of the more (unsuccessful) handovers and decoding errors.

\nosection{Out-of-Order Delay (OOD)\footnote{The OOD of a packet is measured as the time difference between when a packet arrives at the receive buffer and when its previous packets have arrived \cite{chen2013measurement}. It normally does not affect throughput but goodput because most applications require in-order data delivery.}}
As shown in \figref{fig:oood}, we found that BBR has much fewer packets with OOD than CUBIC (\ie 0.80\% versus 5.68\% at 300 \kmh, and 1.53\% versus 5.48\% at 350 \kmh), primarily because of its lower RTT and PLR. Regarding long tail aspect, CUBIC has a much more serious issue: 95\% and 98\% percentile can reach 100 $ms$ and 1 $sec$ respectively, which can significantly affect the QoE. From the mobility level perspective, it only incurs marginal impact in our measurements.

\nosection{Summary of Key Findings}
Our study show that increasing the speed from 300 \kmh to 350 \kmh reduces the TCP goodput by above 40\% and increases the loss rate by up to 92.97\%, while does not significantly affect the RTT; in a high-mobility environment, BBR performs reasonably well by preserving its key advantages (compared to CUBIC) such as being robust to random losses and incurring a smaller amount of bytes-in-flight to potentially mitigate the bufferbloat issue, and thus more efficient in network utilization. 

\begin{figure*}
	\centering
	\subfigure[Size vs. Download Time]{\includegraphics[width=0.33\linewidth]{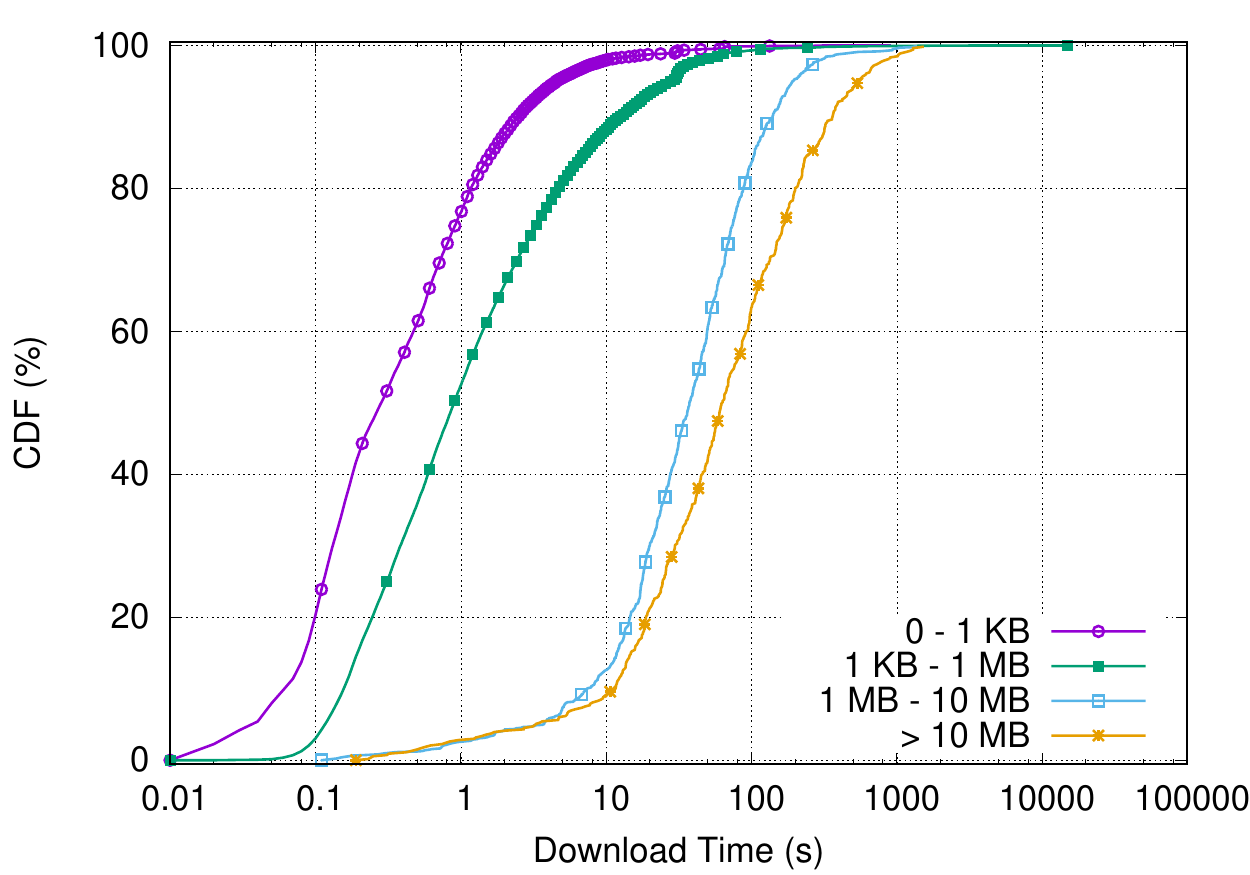}\label{fig:http_size_dur}}
	\subfigure[Completion Percentage]{\includegraphics[width=0.33\linewidth]{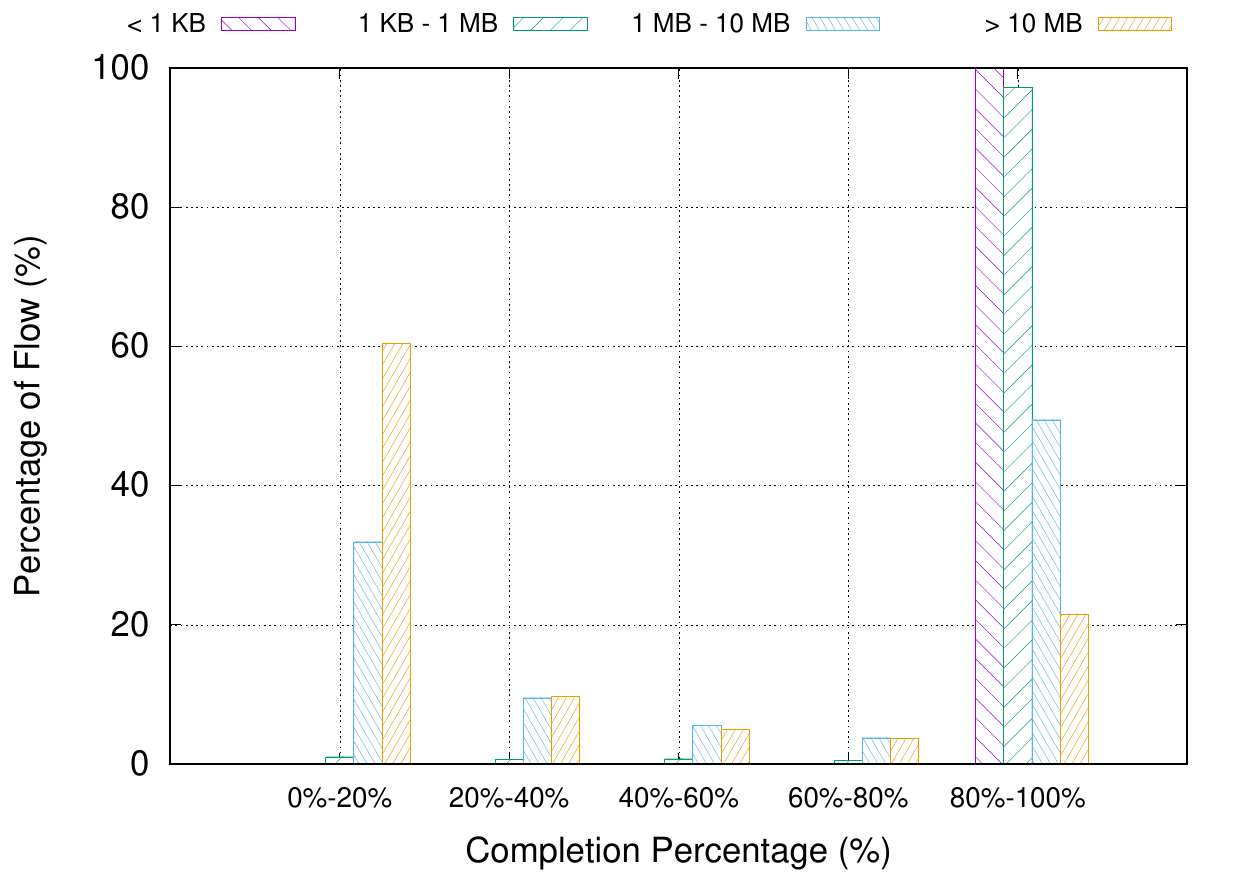}\label{fig:complete_percentage}}
	\subfigure[Time to first byte]{\includegraphics[width=0.33\linewidth]{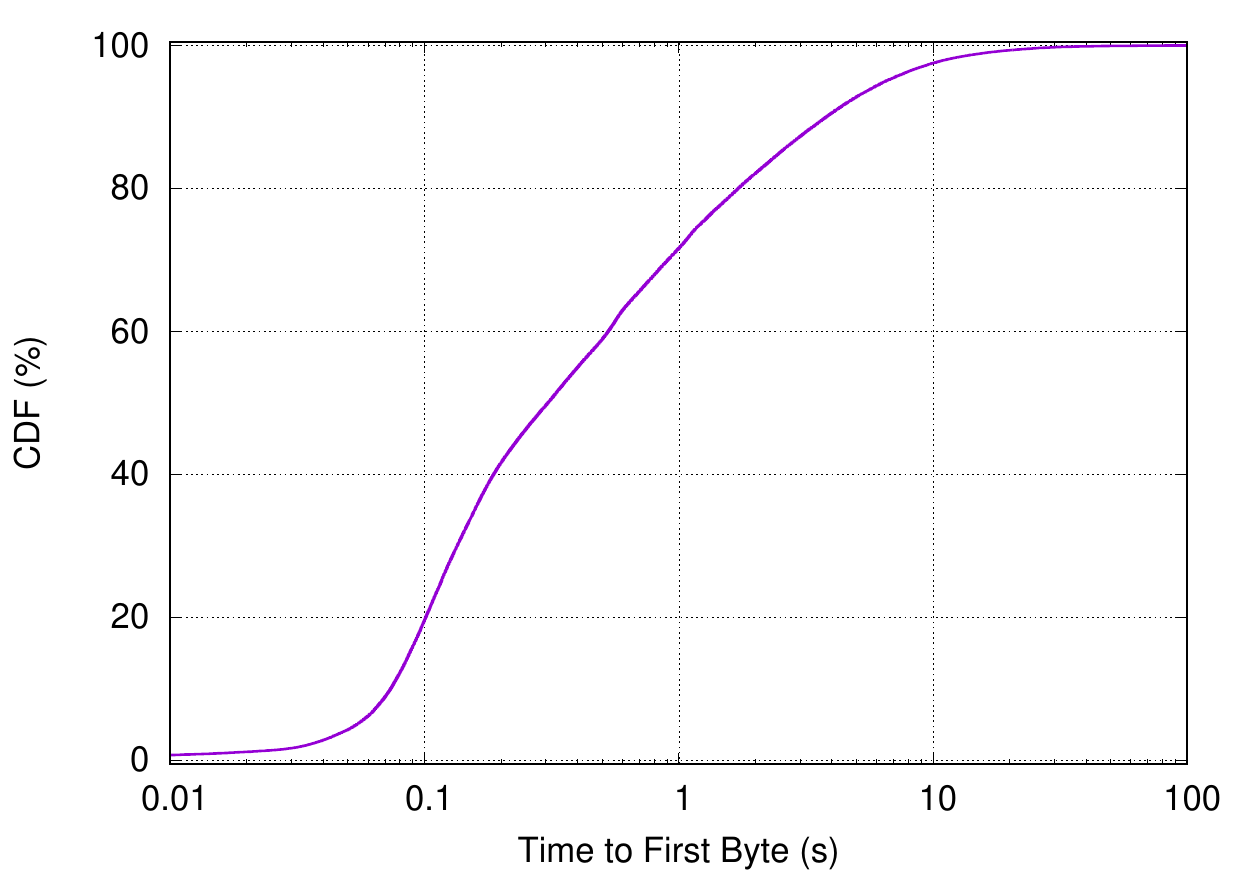}\label{fig:response_time}}
	\figcaption{HTTP QoE Characterization.}
	\label{fig:http_profile}
\end{figure*}

\subsection{User, Flow and Traffic Characteristics}\label{ssec:flowchar}
Understanding the Internet usage pattern of the on-board passengers is vital to network optimization in terms of bandwidth provision and traffic engineering. Given that the train-mounted LTE gateway is providing free WiFi service to all the passengers, it becomes an ideal spot to collect data in the wild.
Inspired by the previous measurement studies of wired \cite{zhang2002characteristics}, WiFi \cite{chen2012network} and LTE \cite{huang2013depth} networks, in this HSR context, we are particularly interested in characterizing the application content, data flow, and the aggregated traffic pattern. For high-level statistics, TCP is still the main transport protocol carrying the Internet traffic from the on-board passengers -- 96.95\% of the data traffic uses TCP, 2.81\% of them uses UDP. Among the TCP flows, 52.94\% and 44.15\% of them are used by HTTP and HTTPS respectively, and the rest are used by other protocols such as SMTP and FTP.

\begin{figure*}
	\centering
	\subfigure[Device/User Concurrency]{\includegraphics[width=0.33\linewidth]{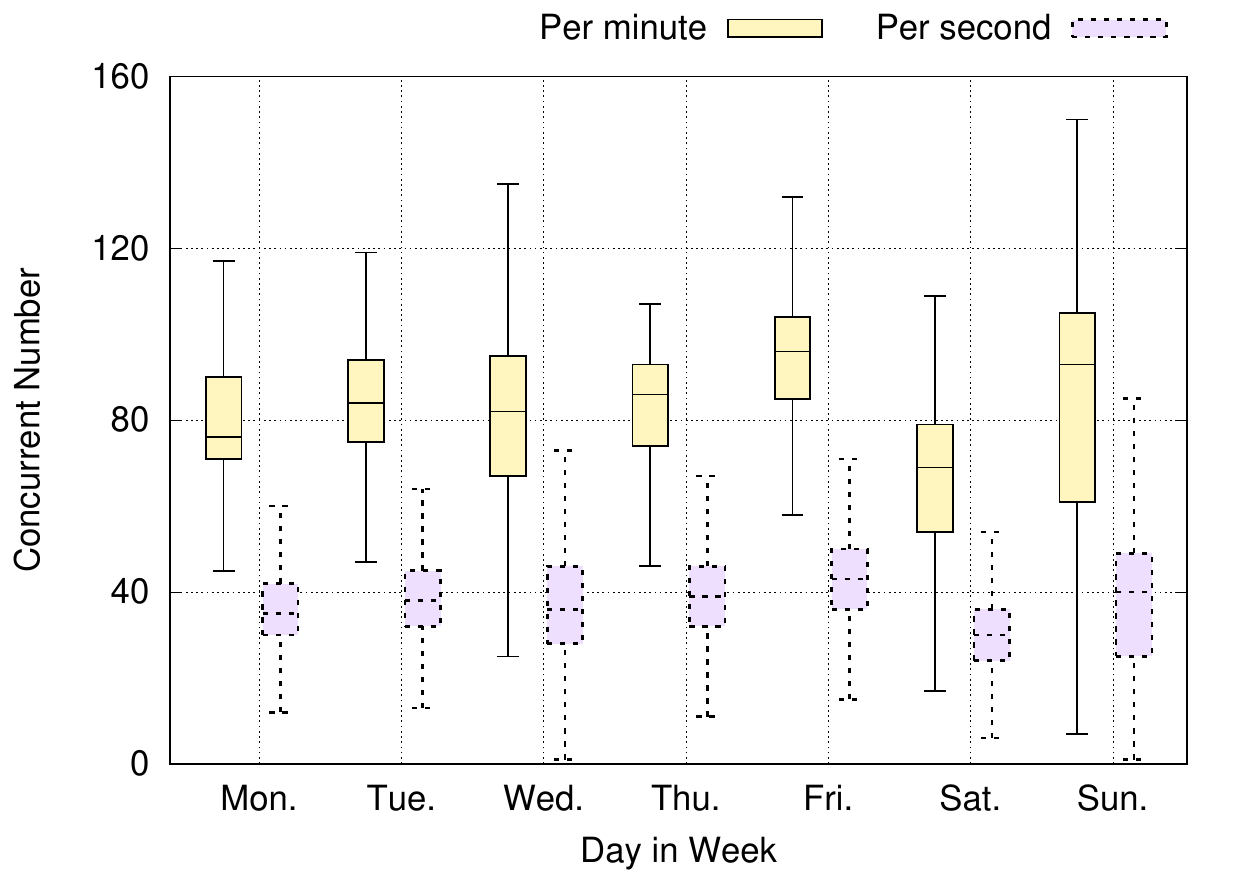}\label{fig:concur_user}}
	\subfigure[Flow Concurrency]{\includegraphics[width=0.33\linewidth]{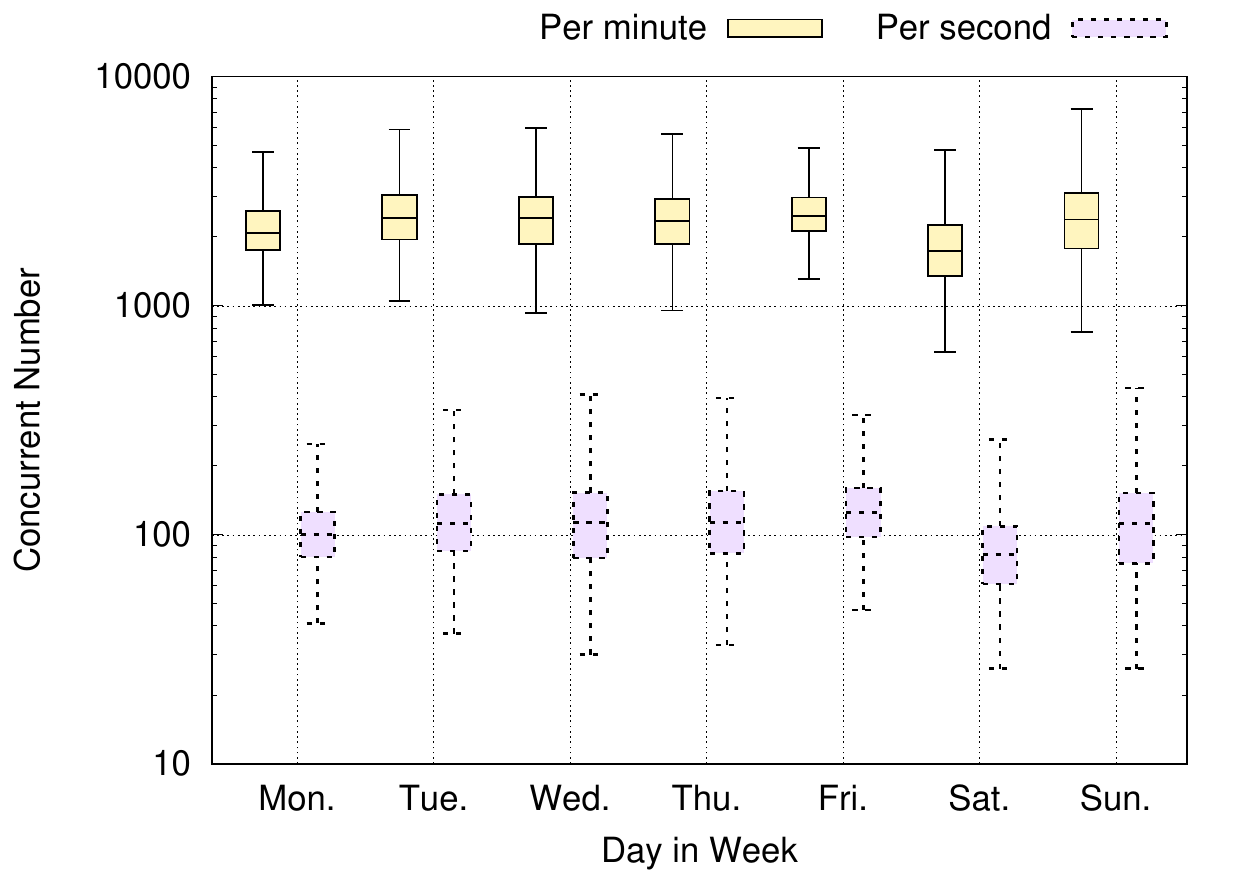}\label{fig:concur_flow}}
	\subfigure[Diurnal Pattern ]{\includegraphics[width=0.33\linewidth]{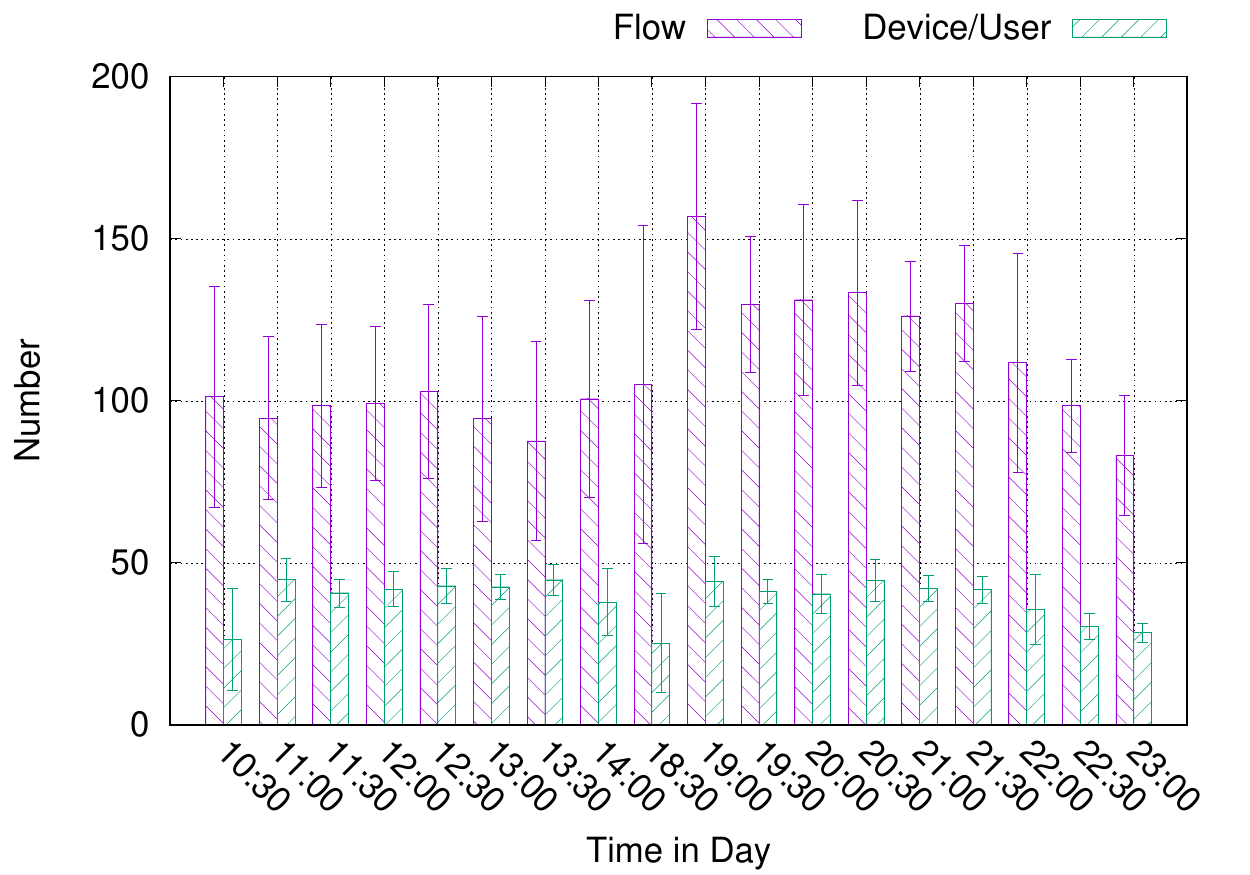}\label{fig:concur_flow_number}}
	\figcaption{User Traffic Pattern.}
	\label{fig:concur_time}
\end{figure*}

\nosection{Web Content Profile}
Given that HTTP(S) data dominates the Internet traffic (94.13\%), we first dive into the high-level statistics based on the content type, regardless of the protocol version. We classify the object type into image, audio, video, application (\eg octet-stream, json, Javascript) and text based on the HTTP header. Note that we did not include the HTTPS traffic due to its encryption nature. From \figref{fig:content_type}, we observe that application data (66.99\%) and video (23.35\%) consume most of the traffic in terms of the data size. Specifically, based on the remote IP address analysis, we discover that the majority of the application data are composed of video, gzip file, WeChat text\footnote{WeChat is a popular Chinese multi-purpose messaging, social media and mobile payment app.}, \etc.
In terms of the number of objects, text (75.51\%) dominates the web usage.
Regarding the object size (\figref{fig:http_flowsize}), the majority (90\%) of the text objects are smaller than 500 bytes; image objects have a larger size range, \ie from 1 KB (10-th percentile) to 40 KB (90-th percentile). Audio and video objects
have median sizes of 34 KB and 125 KB respectively. Despite being the largest, they (in particular, the video objects) are still much smaller than those consumed over typical LTE networks when the smartphone is stationary, based on our controlled experiments. This is likely attributed to the video servers' rate adaptation algorithms or the high abandonment rate due to poor network conditions on HSR to be described shortly.
\figref{fig:http_thp} calculates the per-object data rate, defined as the size divided by the object transfer time (including the HTTP request delay). We only calculate data rate for objects larger than 16 KB to ensure meaningful results.
Overall, the data rates are low -- we barely observe those higher than 1 $Mbps$.
We observe that text has the lowest per-object throughput (median of 233 $Kbps$), followed by image and application.
In addition, we see very low data rate for video objects with a median of only 372 $Kbps$, which likely discourages user for watching videos. The results again indicates the challenges of streaming multimedia contents on HSR.

\nosection{HTTP QoE Characterization}
From our passive data, we are able to infer the QoE of HTTP to some extent, including download time, completion percentage (\ie proportion of received bytes to the object size indicated in the HTTP header), and time-to-first byte (TTFB \cite{halepovic2012can}). Here our analysis focuses on text, image, and application objects given that they dominate the web browsing content (97.53\% in terms of the size).
We first show the CDF of the object size versus download time of the objects with 100\% accomplish rate (\ie those fully downloaded) in \figref{fig:http_size_dur}. While the data rate is fairly low, 80\% of the objects smaller than 1 KB can be finished within 1 second, providing reasonable QoE such as receiving a simple notification message.
On the other hand, although most of the large text/images do not exceed 1 MB, 20\% and 10\% of the download time can be longer than 5 and 10 seconds respectively, significantly hurting the client's user experience. Such high latency is more likely to occur during Type II and III handover.
Second, when the network condition is poor, we often experience an incomplete object transfer due to user abandonment or browser timeout, in particular for large objects.
In \figref{fig:complete_percentage}, we observe that when the object size is small (sub-KB or several KB), the probability that they were fully downloaded are 99.99\% and 96.59\% respectively. However, when the object size is larger than 10 MB, completion percentage of 100\% drops sharply to a surprisingly low value of 16.96\%. 
Finally, we plot the results of TTFB in \figref{fig:response_time}. We observe while the median value is 300 $ms$ which is generally considered as acceptable, the top 25\% and 10\% percentiles can reach higher than 1 and 10 seconds respectively. Overall, the above measurements imply that the QoE on HSR-LTE is far from being satisfactory. 

\nosection{User Traffic Pattern}
It is important to know how many active WiFi users are on the train and how many flows they generate simultaneously, for the purpose of future network infrastructure provisioning and traffic scheduling.
We compute the statistics across different days and hours. From \figref{fig:concur_user}, we observe that Saturday turns out to be day with the least network usage. Our experiences as frequent HSR travelers are that more passengers on that day are traveling with family, and therefore spend more time in talking with others rather than surfing the Internet.
On weekdays or Sunday there tend to be more business travelers.
For the number of concurrent users per second, the median value across different days range from 30 to 43, and the maximum value can reach to up to 98 (out of a typical number of 556 passengers on a fully boarded HSR train). This value is about half of the number when we count per minute. In terms of number of flows (\figref{fig:concur_flow}), the median value is 106 and 2248 per second and per minute respectively. Finally, we report the passengers' diurnal pattern (\ie number of flows and device per second) in \figref{fig:concur_flow_number} at the granularity of half an hour. 
Compared to typical diurnal patterns for residential Internet usage \cite{john2008trends}, the diurnal patterns here are less prominent as attributed to the unique context of passenger traveling by train.
We observe though users are slightly more active in the evening, especially between 19:00 and 21:30 when travelers tend to relax by using Internet for entertainment. Overall, the current on-board WiFi service is only serving less than ten percent of the passengers. 

\subsection{Remarks on Active-Passive Measurements}
While we take both active and passive measurements from the same series of high-speed trains, it is difficult to perform the direct comparison between them. For instance, we observe much lower throughput in passive measurements than active ones (300 $Kbps$ vs. 5 $Mbps$), which can be attributed to the large (HTTP/TCP) protocol overhead for short flows and potential performance bottleneck on either WiFi links, or the inherently deficient design of the multi-tenant antenna system (shared by 9 SIM cards) and the flow-level round-robin flow scheduling mechanism. The high-level statistical results from the active measurements will shed light on the choice of TCP variants and even multi-path transmission mechanism development atop for application-specific server deployment for high mobility data networking in general. 

%% file: cubic.tex
\section{TCP-LTE Analysis in High Mobility}\label{sec:cubic}
From this section, we start to examine the interaction of TCP-LTE performance from a handover-centric view under different mobility level in a finer-grained manner. Specifically, we use CUBIC as a case study in this section, and defer its comparative performance study with BBR in \secref{sec:bbr}.  
\begin{table}[b]
	\tiny
	\begin{tabular}{rlllllllllllllllllllllllll}
		\toprule
		Speed &
		\multicolumn{2}{c}{Throughput} &
		\multicolumn{1}{c}{PHY rate} &						
		\multicolumn{3}{c}{HO Duration (s)} &
		\multicolumn{3}{c}{HO count per 150s-trace} \\
		(\kmh) & All & w/o HO & w/o HO & I & II & III & I & II & III \\
		\midrule
		0 & 13.25 & 13.25 & 14.68 & --- & --- & ---  & --- & --- & --- \\		
		200 & 8.44 & 8.63 & 10.19 & 0.35 & 3.13 & 4.03 & 4.41 & 0.18 & 0.18 \\
		300 & 6.95 & 7.21 & 8.89 & 0.16 & 2.09 & 3.98 & 9.36 & 0.96 & 0.84 \\
		350 & 2.22 & 2.35 & 3.00 & 0.21 & 1.91 & 5.49 & 14.70 & 2.00 & 1.49\\ 
		\bottomrule
	\end{tabular}
	\tabcaption{TCP-LTE performance and handover statistics under different mobility level regarding to no (w/o), successful (I), RLF (II) and NAS (III) handover.} 
	\label{tab:speed}
\end{table}

\begin{figure*}
	\begin{minipage}[b]{0.34\linewidth}
		\includegraphics[width=\linewidth]{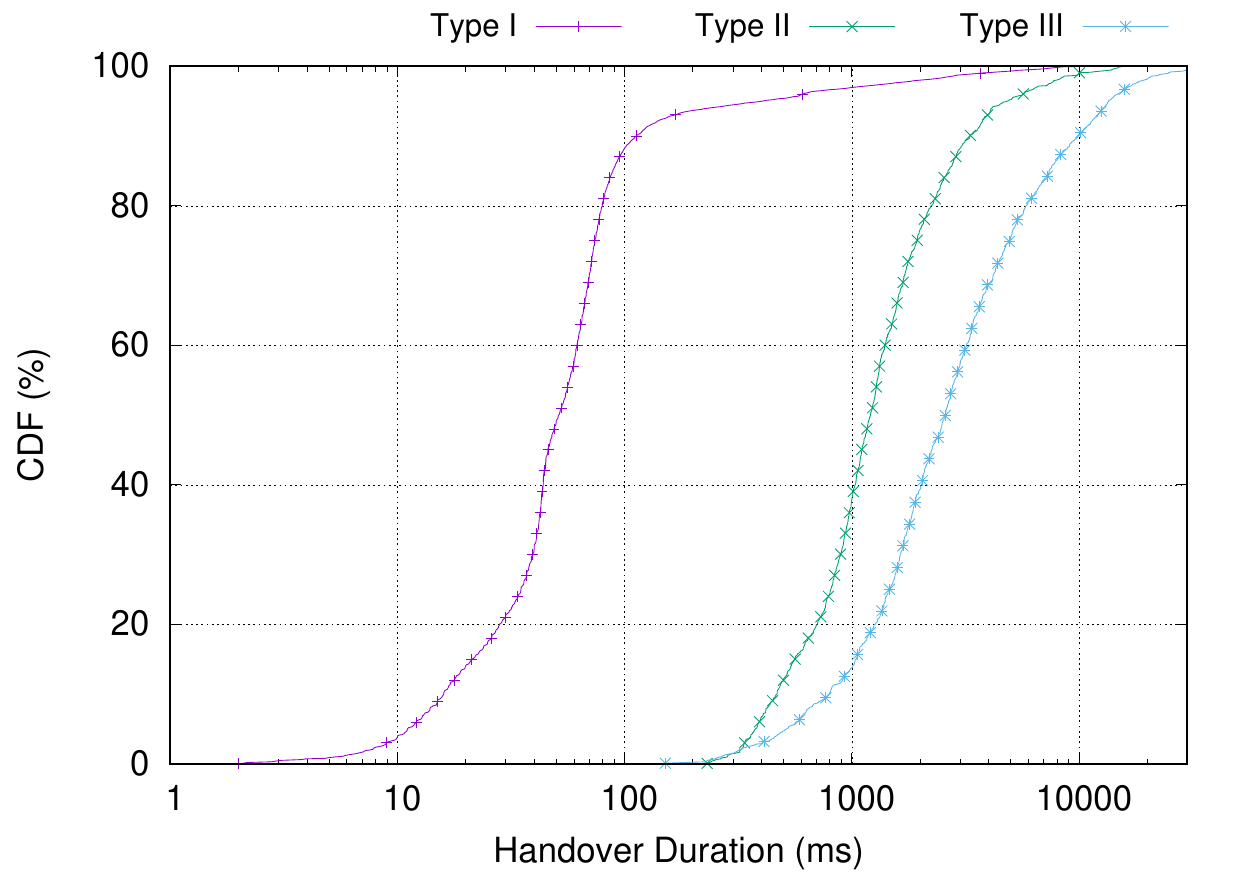}
		\figcaption{Handover Duration.}
		\label{fig:ho_dur}
	\end{minipage}		
	\begin{minipage}[b]{0.655\linewidth}
		\centering	
		\subfigure[Instantaneous impact ]{\includegraphics[width=0.49\linewidth]{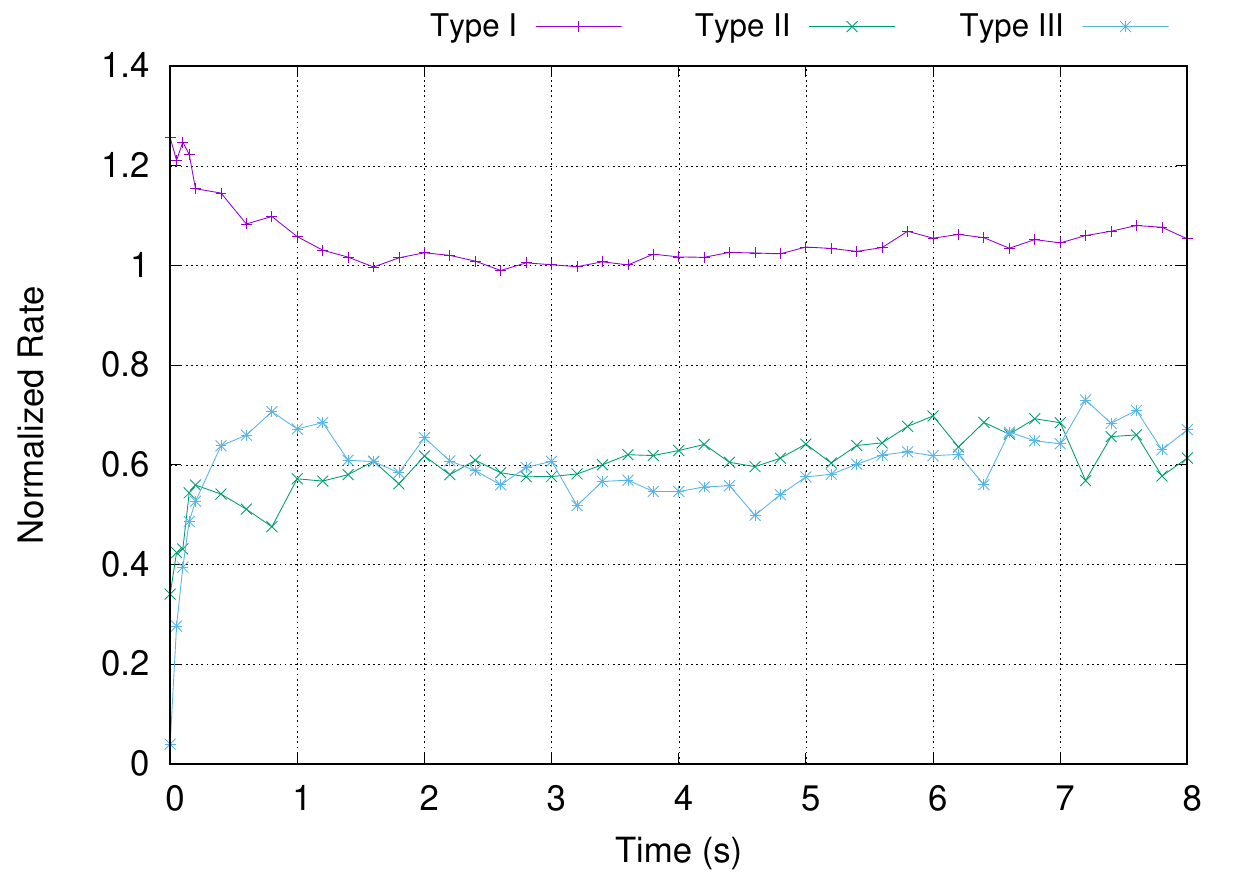}\label{fig:ho_instant_cubic}}
		\subfigure[Near effect]{\includegraphics[width=0.49\linewidth]{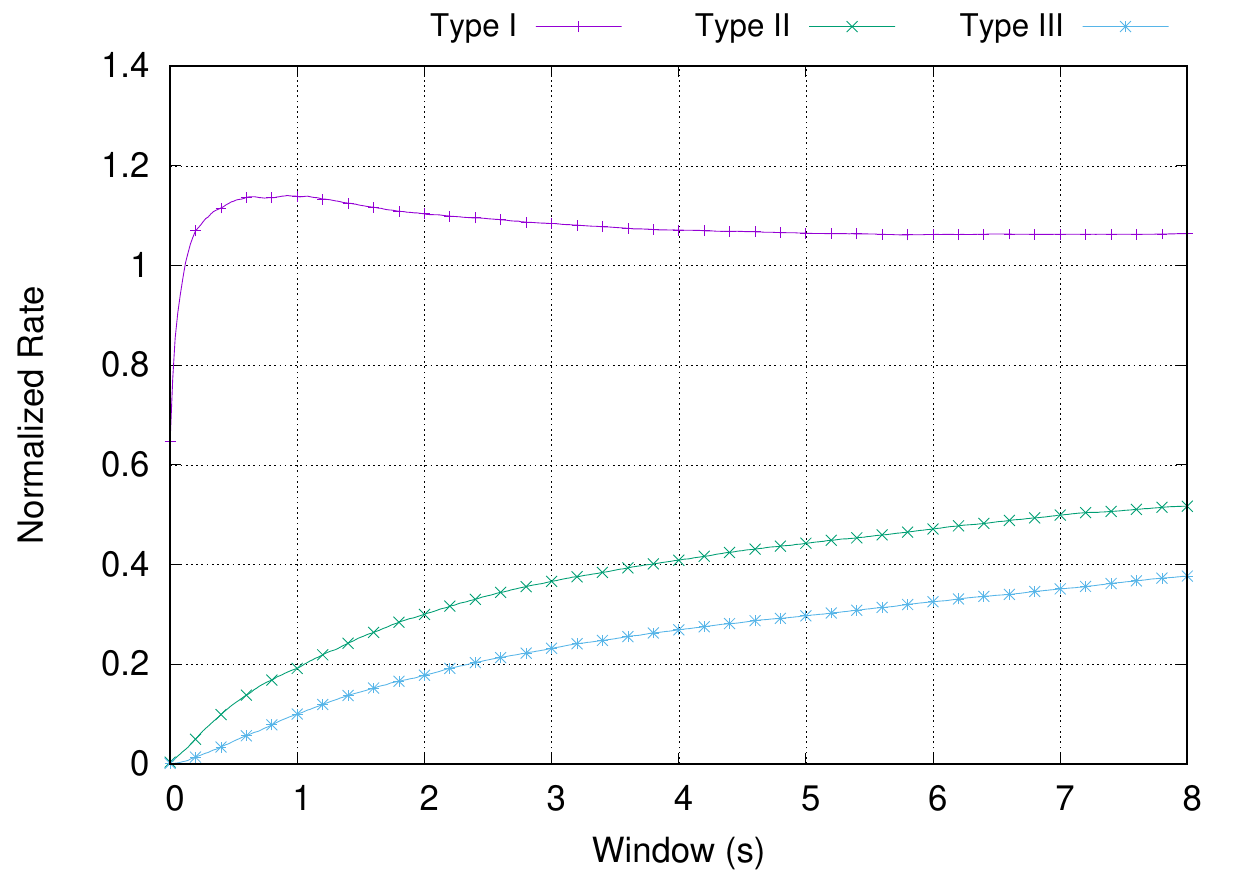}\label{fig:ho_ne_cubic}}
		\figcaption{Handover-centric Analysis of CUBIC.}
		\label{fig:ho_cubic}
	\end{minipage}
\end{figure*}

\subsection{A Mobility-level View}\label{ssec:cubic_mobi}
In our study, mobility level is categorized as stationary, low mobility (200 \kmh) and high mobility (300 and 350 \kmh). Regarding the dataset, it is worthwhile to note that: 
\circled{1} The speed of choice is at which train maintains in most of the data collection time during the whole journey; 
\circled{2} The low mobility (\ie 200 \kmh) data was not common and it was collected over the same route (\ie Beijing-Shanghai) on the date when train happens to travel at that speed after snowing for safety reasons; 
\circled{3} We collect the data in stationary case because its signal has experienced the same path loss (including the carriage penetration loss) and thus it serves as a more appropriate baseline. From \tabref{tab:speed}, we make two key observations \It{as the mobility level increases:}
First, during the period without handover, TCP throughput ($Mbps$) drops from 13.25 to 8.44 (36.3\%), 6.95 (47.6\%) and 2.22 (83.2\%) as the mobility increases from static to low (200 \kmh), high (300 \kmh) and even higher (350 \kmh) respectively \It{in a nonlinear fashion}, which is also reflected in the PHY rate. In fact, they are logically correlated -- the increasing mobility brings severer Doppler spread and channel estimation error, and thus cause higher BER and makes the basestation more conservative in assigning MCS, reducing the number of TB and thus TCP throughput. We note that UDP will be a better choice because it excludes the impact of  congestion control behavior. However, in our experiments we found the carrier will limit the UDP traffic rate (to 1 $Mbps$) from time to time.  
Second, there are more and longer periods of (unsuccessful) handover in the 150-$sec$ traces. Their impact on throughput will be further analyzed later in this section.

In the rest of the paper, we focus on analyzing the 350 \kmh traces as they represent the most challenging HSR networking scenario for today. 

\begin{figure*}
	\begin{minipage}[b]{0.65\linewidth}
		\centering
		\subfigure[Instantaneous Impact]{\includegraphics[width=.495\linewidth]{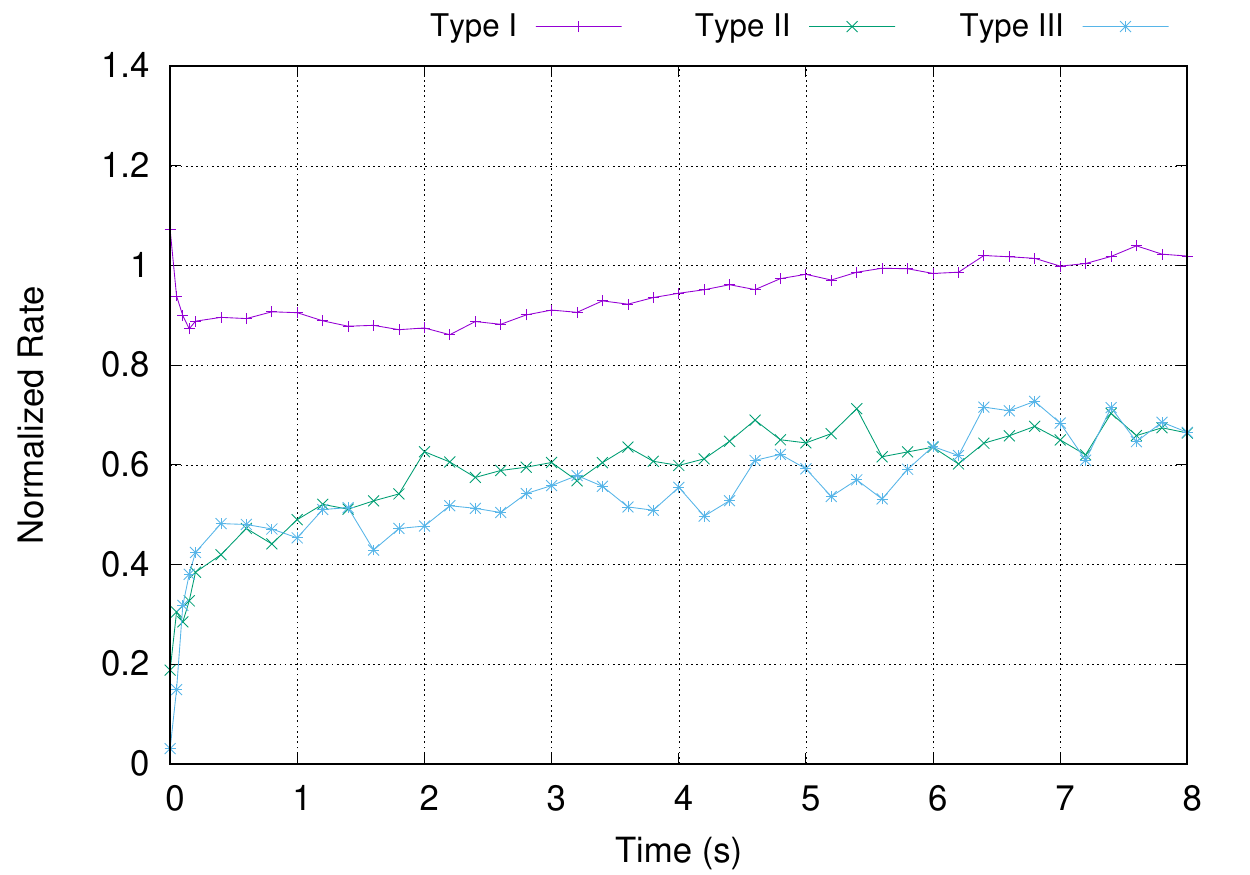}\label{fig:ho_instant_bbr}}
		\subfigure[Near effect]{\includegraphics[width=.495\linewidth]{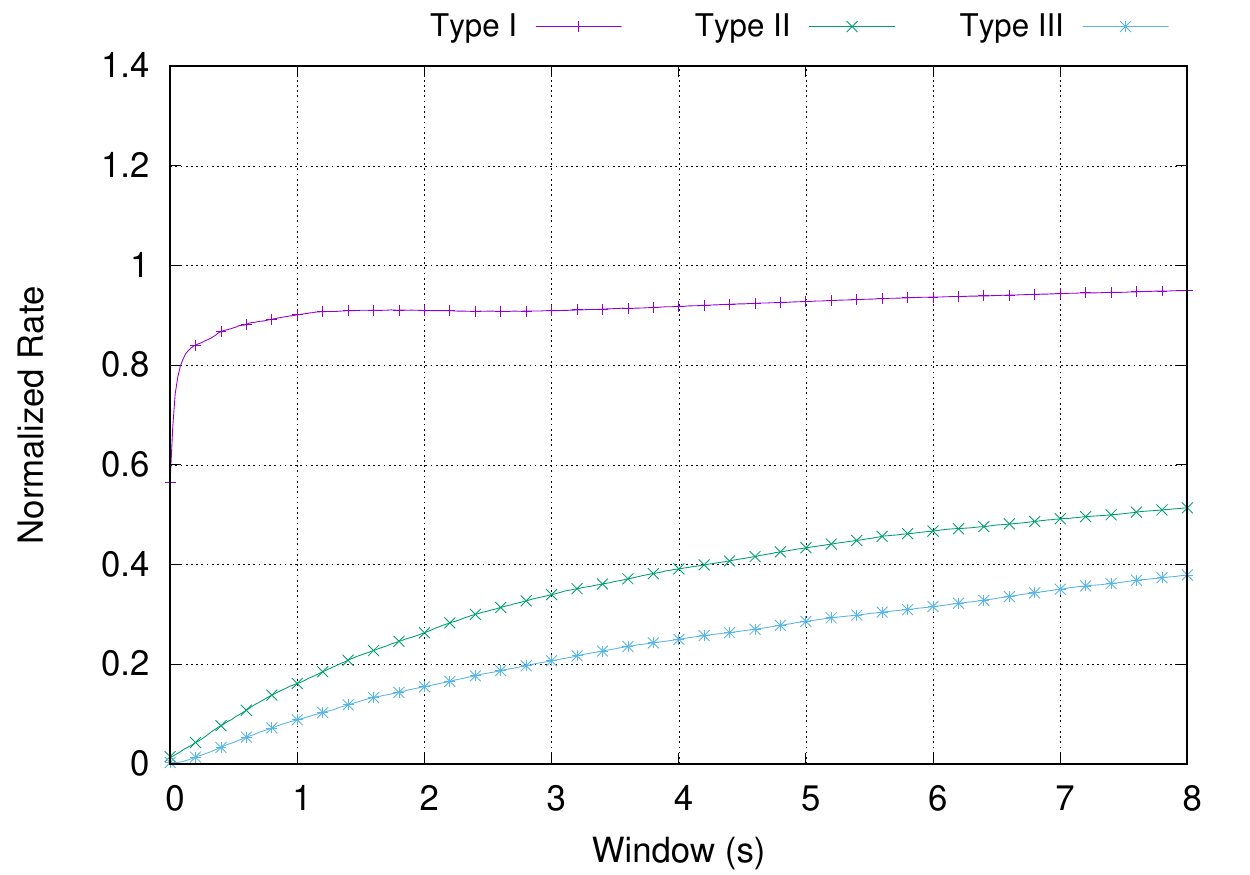}\label{fig:ho_ne_bbr}}
		\figcaption{Handover-centric Analysis of BBR.}
		\label{fig:ho_bbr}
	\end{minipage}
	\begin{minipage}[b]{0.345\linewidth}
		\centering
		\includegraphics[width=\linewidth]{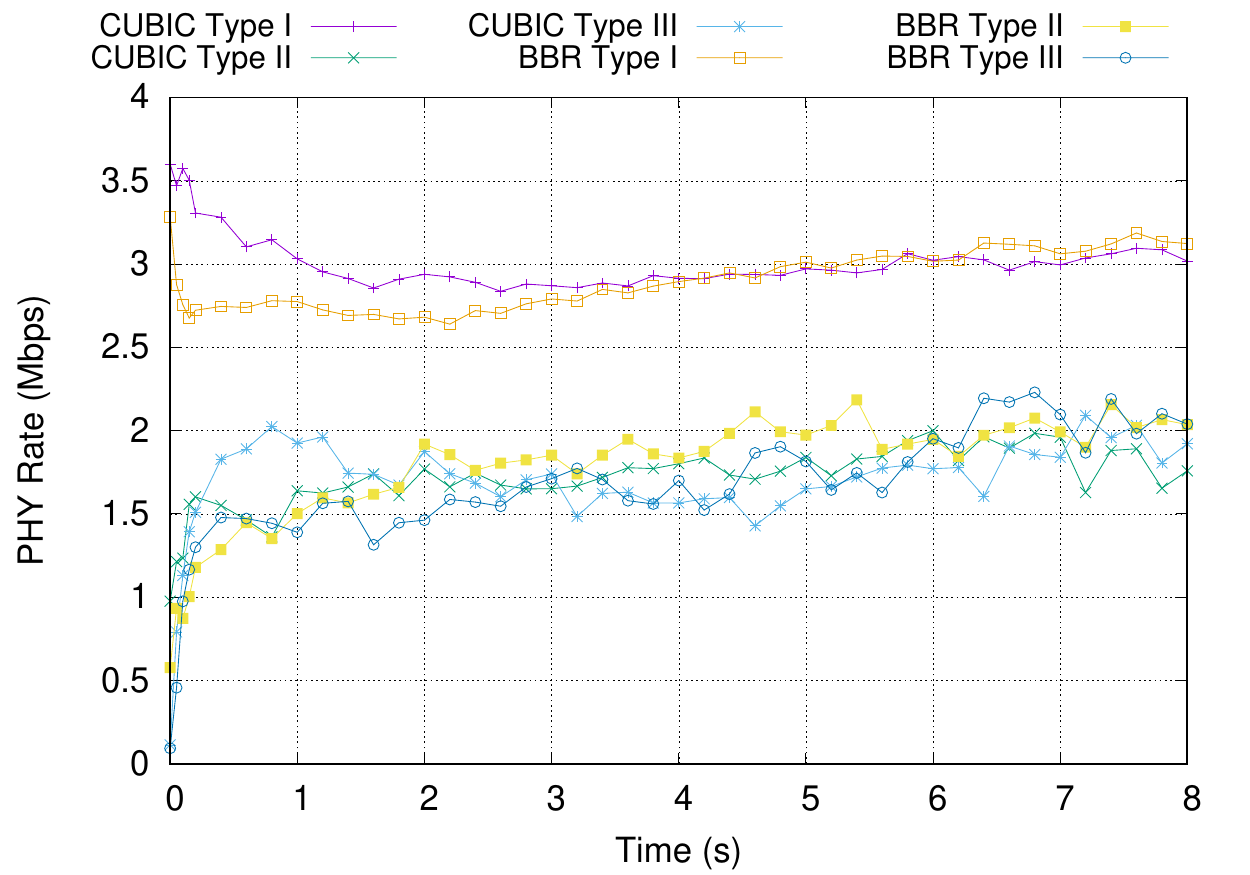}
		\figcaption{Comparative Performance.}
		\label{fig:ho_comp}
	\end{minipage}	
\end{figure*}

\subsection{A Handover-centric View}\label{ssec:cubic_ho}
Handover can cause different level of TCP disruptions, depending on how long it takes and whether it is successful. 
As shown in \figref{fig:ho_dur}, 85\% of the type I handover finishes within 100 $ms$, which has a high chance to be hidden from the TCP as their RTTs are often more than that period (\figref{fig:rtt}).
On the other side, more than half of the type II/III handover last more than 1 second, and the top 25\% of type II and III handover are even longer than 2 and 5 seconds respectively. To quantify their negative impact, we first study how does the data rate change after the handover. We denote \It{Normalized Rate} as the PHY rate during the time interval (\ie 200 $ms$) divided by the average PHY rate among all traces to even out the different networking performance across the traces. Specifically, handover is shown as a single point in the origin and its interval is regarded as $HO_{\textrm{start}}$ to $HO_{\textrm{end}}$ when in calculation.     
Note that the reason we choose PHY rate instead of TCP throughput is that there exists tens of $ms$ delay between the on-chip time of LTE protocol message (reported by \Mod{MobileInsight}) and the system time of TCP pcap trace. Hence, the handover and PHY information both reported from \Mod{MobileInsight} should be better aligned in timing. 

\nosection{Instantaneous Impact} One way to quantify the impact of handover is to simply observe the instantaneous data rate after it happens, or $HO_{\textrm{end}}$. Taking CUBIC as an example, in \figref{fig:ho_instant_cubic}, we make three key observations: 
First, for type I handover, the normalized rate is almost unaffected by handover and reaches the average rate immediately after the handover, indicating only a few packets are delayed but connection is marginally affected. This is because most type I handover completes within 100 $ms$, which will unlikely trigger RTOs. In fact, we even observe tens of $ms$ data burst right after the handover ends. This is because the lossless nature of type I handover will ensure the data will not get lost during the handover procedure and can be delivered to the UE from the target cell instead of the server. Note that the normalized rate can reach above 1 since it is normalized by the average rate over the trace, which can certainly be smaller than some instantaneous rate out of the handover period.
Second, type II/III handover has a more significant impact on the throughput -- it often turns the connection down for a longer time (\ie longer than 2 seconds in more than quarter of the time) and is more likely to trigger RTO and slow start. 
Third, although both type II and type III handover are triggered by radio link failure, the data rate at time 0 of the former one is typically higher. This is because type II handover (by definition) is able to transfer the UE context as well as the buffered data to the target cell, while type III handover fails to do and needs upper-layer retransmission.

\nosection{Near Effect} In practice, the (negative) impact of handover is beyond the instantaneous rate, \ie user experience is affected not only after the handover ends, but also during its period. We define such phenomenon as \It{Near Effect}, and denote \It{window} of length $x$ where $x$ means $[HO_{\textrm{start}}, HO_{\textrm{end}}+x]$. We quantify the near effect by computing the ratio of the average PHY rate of the \It{handover incorporating that window $x$} to the average PHY rate among all traces.
From \figref{fig:ho_ne_cubic}, the key observation we make is that while type I handover shows a similar pattern as it shows in the instantaneous impact because of its short duration, type II/III handover has a much lower normalized rate in terms of near effect, primarily because they themselves have a longer (handover) duration, together with the higher probability of multiplicative decrease and slow start due to packet loss and RTO respectively. Specifically, it does not reach half of the average rate after 10 seconds for type III handover. 

%% file: bbr.tex
\section{Comparing BBR with CUBIC}\label{sec:bbr}

\begin{figure*}
	\begin{minipage}[b]{0.655\linewidth}
		\centering
		\subfigure[Round-trip propagation time]{\includegraphics[width=.49\linewidth]{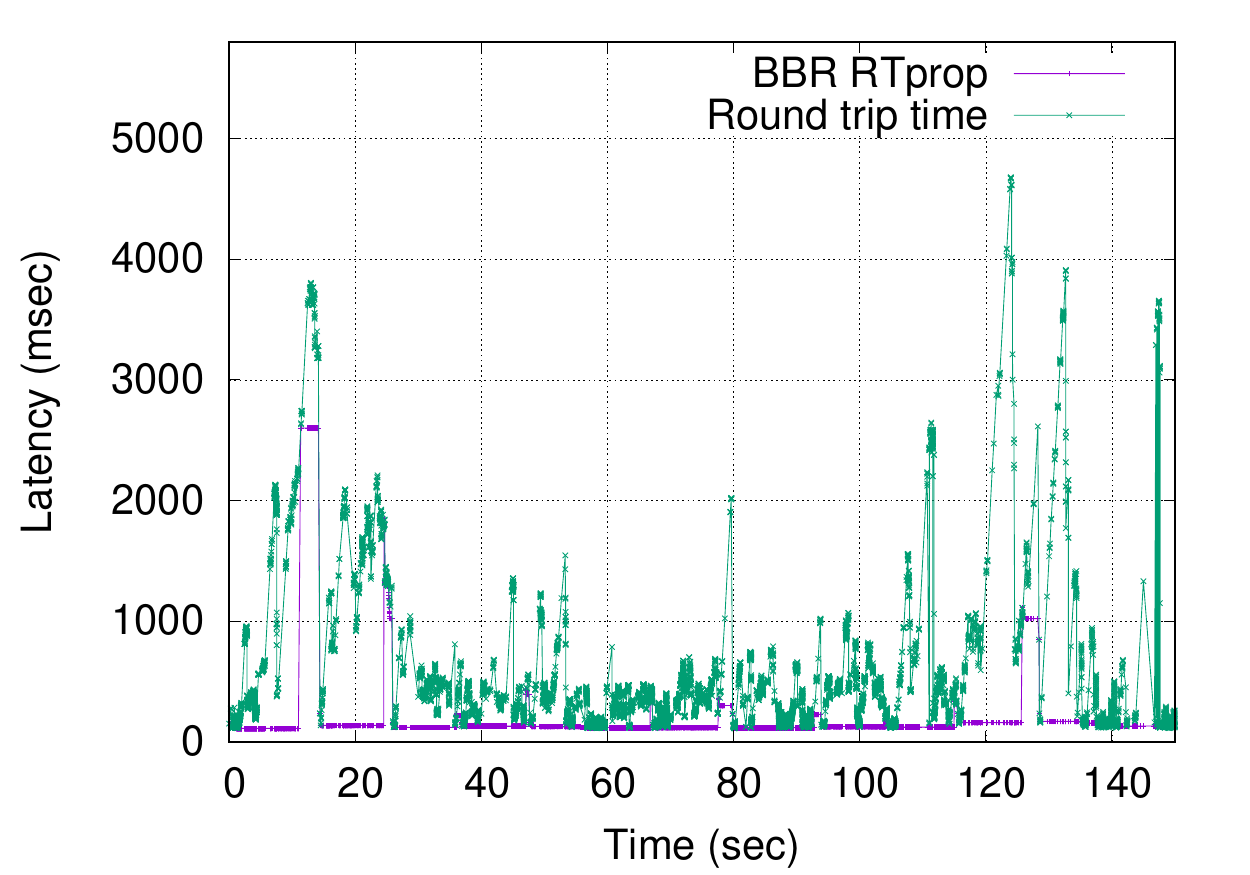}\label{fig:bbr_rtprop}}
		\subfigure[Bottleneck bandwidth]{\includegraphics[width=.49\linewidth]{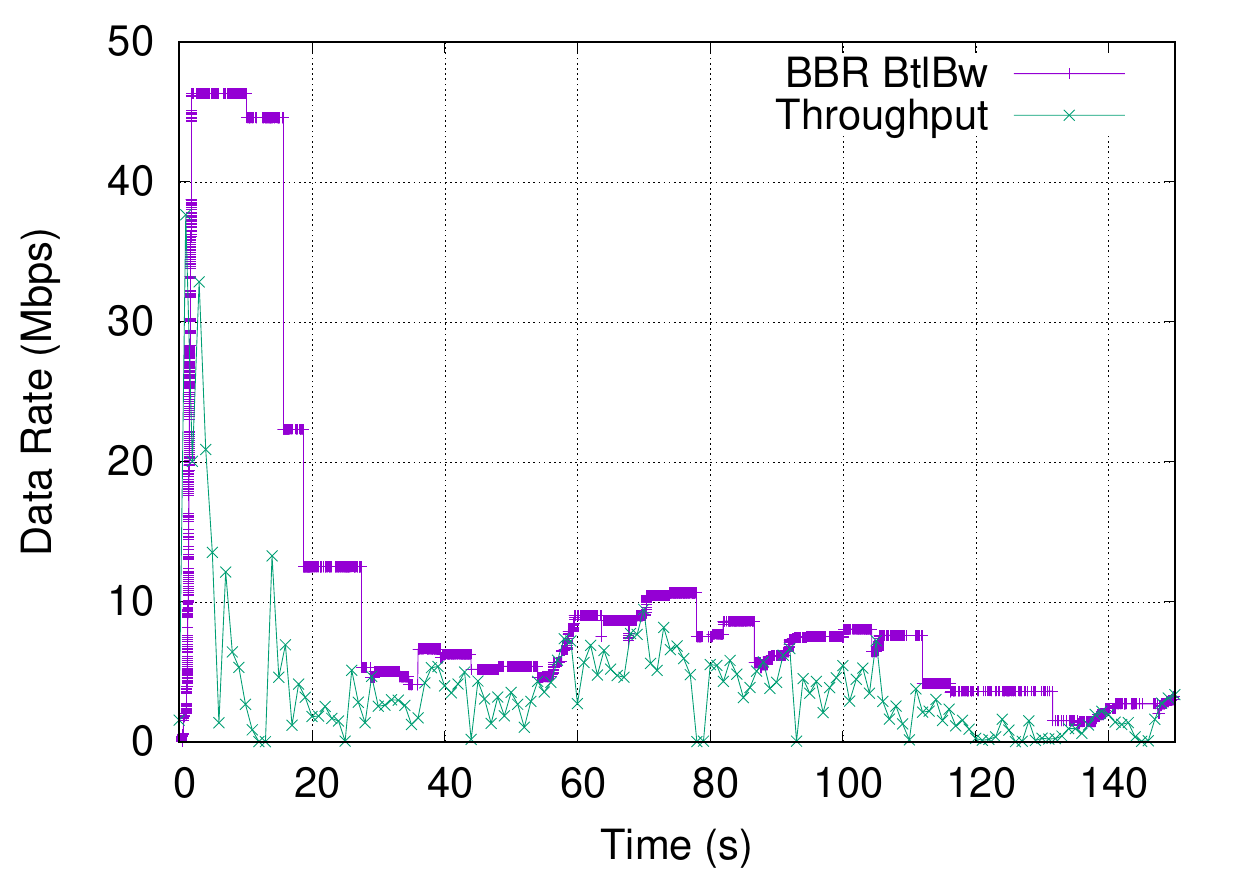}\label{fig:bbr_btlbw}}
		\figcaption{BBR's Parameter Trace.} 
		\label{fig:bbr_para}
	\end{minipage}
	\begin{minipage}[b]{0.34\linewidth}
		\centering
		\includegraphics[width=\linewidth]{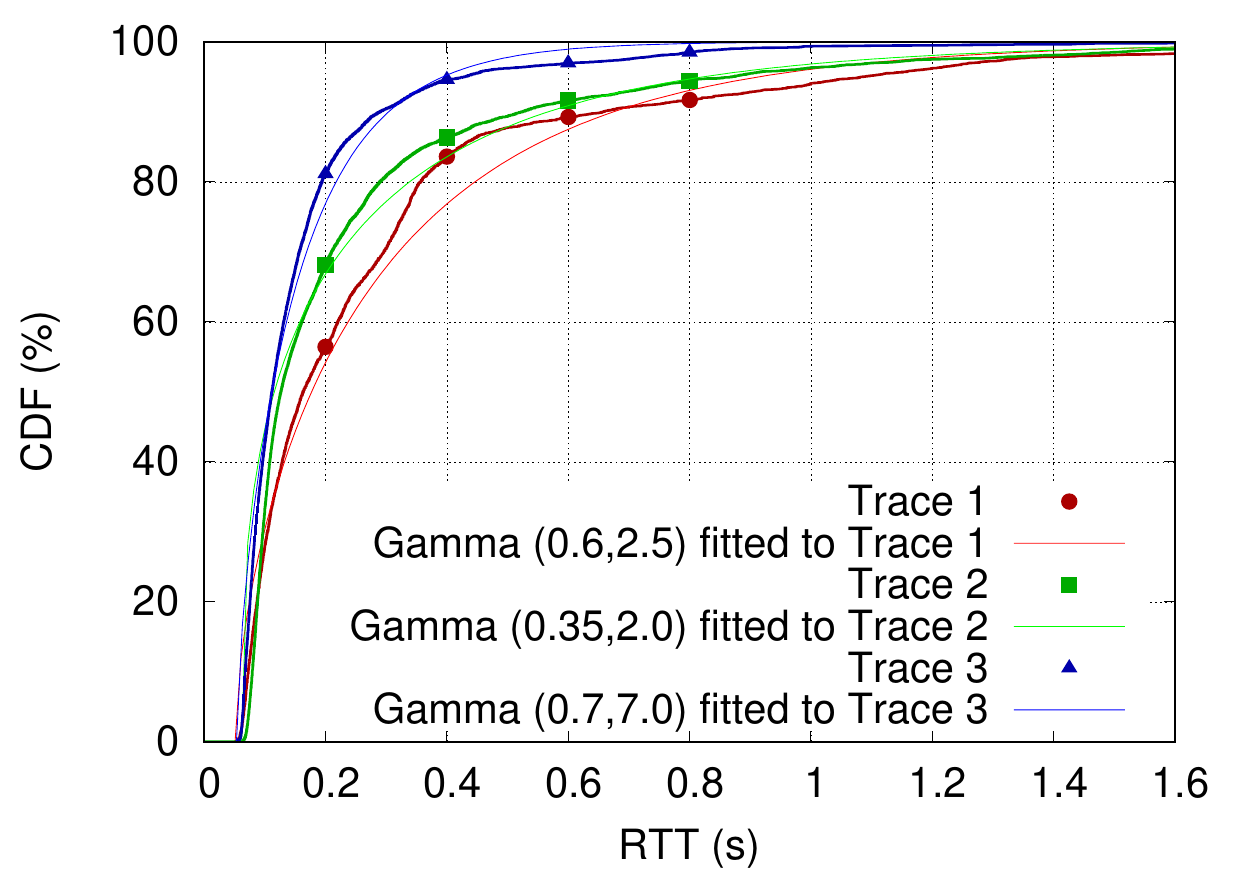}
		\figcaption{RTT Distribution.}
		\label{fig:rtt_pdf}
	\end{minipage}	
\end{figure*}

\subsection{BBR is More Handover-Agnostic}\label{ssec:bbr_ho}
We follow the same analysis procedure (in \secref{ssec:cubic_ho}) and study BBR in a handover-centric manner. From the instantaneous impact perspective (\figref{fig:ho_instant_bbr}), the key observation we make is that, for all types of handover, BBR maintains a smoother and slower data rate change than CUBIC. This is because BBR has a intrinsically less radical strategy than CUBIC in expanding its $\mathit{cwnd}$ -- approximately at most 25\% increase for $\mathit{BtlBw}$ at per 8$\cdot \mathit{RTprop}$, which means the same amount of increase in $\mathit{cwnd}$ (and the sent data) is not as much as CUBIC's over the same time window. Note that the lossless nature of type I/II handover is observed in the BBR case as well. 
In terms of near effect (\figref{fig:ho_ne_bbr}), we observe similar pattern as shown in the instantaneous impact.
Finally, by putting CUBIC and BBR in parallel and showing their PHY rate (\figref{fig:ho_comp}), we observe that CUBIC slightly outperforms BBR for all type of handover in the early stage. However, BBR starts to surpass CUBIC after 3 seconds in all cases because it paces smoothly regardless of (random) packet loss. 

\nosection{Remarks}
We would like to point out that type III handover for BBR could potentially represent the worst situation in our study: after the long network disconnection (\ie exceeds 10 $\cdot \mathit{RTprop}$ with certain probability), BBR has to recover from the $\mathit{BtlBw}$ of nearly 0, which is a much smaller basis than type I/II in expanding its $cwnd$. We observe such phenomenon (\eg the normalized instantaneous rate after type III handover cannot even recover to 0.25 after 8 seconds) in a few collected traces, which is however not reflected in statistics (\figref{fig:ho_instant_bbr}) due to their relatively small proportion. 

\subsection{BBR is More Carrier-Agnostic}\label{ssec:bbr_perf}
Recall that in 150-$sec$ trace of 350 \kmh (\figref{fig:goodput_350}), BBR has comparable goodput ($Mbps$) with CUBIC over \ca (5.12 vs. 5.40), but outperforms around 70\% than CUBIC (3.54 vs. 2.06) over \cb. 
Intuitively, CUBIC could suffer over a lossy connection, while BBR is insensitive to packet loss. To verify the fact that BBR outperforms CUBIC in \cb is because the cellular infrastructure causes higher random loss, we carry a concurrent test in both static and 350 \kmh using the same setup as the controlled experiment described in \secref{ssec:ctrl_exp}, except that each TCP packet only carries 1 byte of data and sends at a stable rate of 20 packets per second to avoid self-inflicted congestion. The results summarized in \tabref{tab:rand_loss} show that while \cb does cause higher packet loss in non-congestion conditions, potentially due to its poorer mobility support, less network coverage, \etc, which explains the relatively poor CUBIC performance over \cb. 

\begin{table}[H]
	\centering\small
	\begin{minipage}{\linewidth}
		\begin{tabularx}{\linewidth}{|l|X|X|X|X|} \hline 
			Flow setup & Static A & HSR A & Static B & HSR B \\ \hline\hline
			Loss rate & 0.008\% & 0.21\% &   0.008\% &  1.35\%  \\ \hline			
		\end{tabularx}
	\end{minipage}
	\tabcaption{Random loss rate for concurrent flows.}
	\label{tab:rand_loss}
\end{table}

\subsection{BBR is Suboptimal on High-speed Rails}\label{ssec:bbr_suboptimal}
Our measurements reveal that BBR is able to maintain its desired property of low RTT on HSR, but shows only comparable throughput with CUBIC, not as good as it performs in large-scale WAN -- throughput gain over CUBIC by 2-20x \cite{cardwell2017bbr}. 
Given its model-based nature, it determines its congestion control window size based on its estimation of $RTprop$ and $BtlBw$. Hence, to understand how well the model works, we randomly pick a flow over \ca and plot the time series of these two parameters along with the measured RTT and throughput in \figref{fig:bbr_para}. From \figref{fig:bbr_rtprop}, we can see that BBR's estimated $RTprop$ significantly deviates (0.35x on average) from the RTTs, which can potentially in turn self-throttles its throughput in such networking environment with high RTT variation. On the other hand, even though the estimated $BtlBw$ is generally larger than instantaneous throughput (1.6x on average) (\figref{fig:bbr_btlbw}), the resulting estimated congestion control window (\ie the product of $RTprop$ and $BtlBw$) can still be potentially increased, maybe at a acceptable cost of RTT. Specifically, we argue that the estimation of $RTprop$, \ie the minimum RTT over the last 10 seconds, leads to the underutilization of the connection capacity. We extend this discussion and present a renovated BBR design with experimental evaluation in \secref{sec:improve}.  

%% file: improve.tex
\section{Improving Data Transfer on HSR}\label{sec:improve}
Although our study reveal that TCP performance on HSR is greatly constrained by the imperfect lower-layer coordination, it is always worthwhile revisiting the TCP (rather than cross-layer) design when considering deployment cost. In comparison to CUBIC, BBR achieves similar throughput but much lower RTT (\secref{ssec:tcpperf}), and thus sets a better basis for low-latency networking to provide better QoE for most today's applications. Our goal here is to \It{renovate BBR to further improve the throughput at an reasonable cost of latency.}

\subsection{BBR+ Design}\label{ssec:bbrp}
The design of BBR is intrinsically suboptimal in networking environment where both bandwidth and RTT change rapidly, \eg HSR in our study, because both its bandwidth probing strategy and round-trip propagation time estimation do not adapt to the network dynamics in a agile way. Herein, we improve the BBR design in the following two aspects.

\nosection{Cycling $\boldsymbol{pacing\_gain}$}
Recall that in \secref{ssec:tcp_review}, default cycling $\mathit{pacing\_gain}$ sequence makes BBR pace the sending rate according to the drain rate (\ie $\mathit{BtlBw}$) to keep the bottleneck buffer nearly empty in most of the time. This mechanism is designed for a connection with relatively stable $\mathit{BtlBw}$, not the case in our scenario. Therefore, we adjust the sequence to be more radical to adapt to the HSR environment:
$$ 3/2, 1/2, 3/2, 1/2, \dots $$

\nosection{$\boldsymbol{RTprop}$}
As discussed in \secref{ssec:bbr_suboptimal}, the $\mathit{RTprop}$ BBR estimates is too conservative in HSR scenario when it \emph{has to} be regarded as a constant over the last 10 seconds since the theoretical foundation of BBR is the local stability of $\mathit{RTprop}$ and $\mathit{BtlBw}$. Our intuition is that $\mathit{RTprop}$ needs a compensation term accounting for the network dynamics and we observe that the RTTs in each trace approximately follow a \emph{shifted gamma distribution} (shifted by $\min RTT$) with a fat tail. 
In \figref{fig:rtt_pdf}, we show the randomly chosen three BBR traces (with disparate mean and variance of RTTs) along with their corresponding fitted shifted gamma distributions with different parameters, denoted by $\mathrm{Gamma}(\alpha, \beta)$. For a random variable following a shifted gamma distribution, we have:
$$
\mathrm{E}(\boldsymbol{X}) = \min{X} +  \sqrt{\alpha \mathrm{Var}(\boldsymbol{X})},
$$
where $\alpha$ is the shape parameter of gamma distribution. Therefore, $\sqrt{\alpha\mathrm{Var}(RTT)}$ becomes the natural compensation term to let us get closer to the expectation of actual $\mathit{RTprop}$. Hence, we have:
$$ 
\mathit{RTprop} = \min RTT + \lambda \sqrt{\mathrm{Var}(RTT)}
$$
Here, $\mathrm{Var}(\boldsymbol{X}) = E(\boldsymbol{X}^2) - E(\boldsymbol{X})^2$, and $\lambda = \sqrt{\alpha}$ is a tunable parameter which allows us to trade off between bandwidth and RTT. Specifically, we estimate $E(\boldsymbol{X})$ by using EWMA (\ie Exponentially Weighted Moving Average) with \emph{time-based weight decay} instead of classical EWMA so that the assigned exponentially decayed weights are irrelevant to the sending rate but determined by the time elapsed:
$$
E(\boldsymbol{X}) \approx \frac{1}{c} \int_{-\infty}^{t_0} \mathrm{e}^{-c(t_0-t)} X(t) \mathrm{d} t
$$
Note that the two terms in the new $\mathit{RTprop}$ estimation are complementary to some extent: the former term monitors the long-term bottleneck buffer occupancy, while the compensation term captures the short-term network dynamics.

\begin{table}
	\centering\scriptsize
	\begin{minipage}{\linewidth}
		\begin{tabularx}{\linewidth}{|l|X|X|X|} \hline 
			Congestion Control & BBR & $\textrm{BBR+}_{(\lambda=1/8)}$ & $\textrm{BBR+}_{(\lambda=1/2)}$ \\ \hline\hline
			Goodput (Mean $\pm$ Std in $\mathit{Mbps}$) & 4.44 $\pm$ 3.09 & 5.49 $\pm$ 3.26 & 6.06 $\pm$ 3.25 \\ \hline	
			RTT (25/50/75th percentile in $\mathit{ms}$) & 76/149/311 & 80/242/694 & 98/333/1090 \\ \hline 
		\end{tabularx}
	\end{minipage}
	\tabcaption{Performance Comparison of BBR and BBR+.} 
	\label{tab:bbrp}
\end{table}

\subsection{Evaluation}
We evaluate BBR+ in the same experimental setting in \secref{ssec:ctrl_exp}, except that the servers run BBR, $\textrm{BBR+}_{(\lambda=1/8)}$ and $\textrm{BBR+}_{(\lambda=1/2)}$ respectively, where $\lambda$ is the constant used for calculation of $\mathit{RTprop}$. 
The experimental results summarized in \tabref{tab:bbrp} show that $\lambda$ is effectively controlling the amount of packets to be filled into the (bottleneck) buffers for throughput gain at the cost of RTT -- as $\lambda$ increase from $1/8$ to $1/2$, its throughput gain over BBR also increases from 24\% to 36\%, with increased median RTT of 93 $ms$ and 184 $ms$ respectively, which is still much less than CUBIC. 

\nosection{Remarks} The efforts in this pilot study mean to be inspirational rather than comprehensive. We demonstrate that BBR+ can potentially achieve throughput gain in the applications with a tolerable latency bound. Although the choice of ${pacing\_gain}$ and $\lambda$ in our experiments shows its efficacy, we note that they do not prove to be optimal either theoretically or experimentally. However, we believe there is great potential in the design space of HSR networking. 

%% file: discuss.tex
\section{Discussion}\label{sec:discuss}
\nosection{TCP Variants}
We note there are many alternative (single path) end-to-end TCP variants in the wild, which can be categorized into loss-based \cite{jacobson1988congestion,xu2004binary} and delay-based \cite{brakmo1995tcp,mascolo2001tcp,tan2006compound} congestion control algorithm in general. 
We choose CUBIC and BBR in our study because they both not only have large-scale real world deployments, but also represent the state-of-art solution in each category -- CUBIC provides the best goodput over high-BDP networks \cite{alrshah2014comparative}, 
and BBR in a sense can be regarded as a delay-based approach as it also aims to keep the delay short and even outperforms CUBIC by 2 to 25x in WAN environments \cite{cardwell2017bbr}.
Meanwhile, we are aware that there are recent designs dedicated for cellular access \cite{jiang2012tackling,winstein2013stochastic,zaki2015adaptive,lu2015cqic,xie2017accelerating,leong2017tcp,park2018exll}.
We leave a comprehensive study for future work. 

\nosection{Railway Route}
HSR networking performance may be dramatically different on different routes. Our analysis so far is based on the data collected from Beijing-Shanghai HSR route, which has the best (LTE) coverage among all the routes in China. For other routes with poorer LTE coverage in terms of weaker signal strength and higher packet loss rate, we expect BBR will outperform CUBIC in such cases.

\nosection{Beyond 350 \It{km/h}} 
Recent studies \cite{rula2016ips,rula2018mile} have started to look into the networking performance on airplanes (\ie 800+ \kmh). We believe these two extreme mobility use cases together will call for attention on  making best use of cellular and satellite links for improving efficiency and robustness. 

%% file: related.tex
\section{Related Work}
\nosection{TCP Measurement Study on HSRs (300+ \It{km/h})}
Most prior measurement work on HSR only focuses on the TCP level. 
The study in \cite{jang20093g} shown that ACK compression is common and that spurious retransmission represent more than 50\% retransmission. 
The work \cite{xiao2014tcp} presented the first public large-scale empirical study on TCP performance in HSR scenarios. The main observation is that the TCP throughput is much worse (3x and 2x) than static and driving scenarios, primarily because of the larger RTT jitter and variance, induced by the channel loss and handover.  
Most recently, Li et al. \cite{li2017longitudinal} quantify TCP's poor adaptation to high mobility environments, such as high spurious RTO rate, aggressive congestion window reduction, a long delay of connection establishment and closure, and transmission interruption. In \cite{li2018measurement}, they further discovered a MPTCP with coupled congestion control over multiple cellular carrier setup provides better performance than TCP in the poorer of the two paths, while performs worse than TCP in the better path most of the time. 
Our work differ from them in that we not only look into the LTE protocol message including L1/2 for investigating the root cause of TCP (abnormal) behavior, but also extend it to a comparative study on TCP variants (\ie CUBIC and BBR) to shed light on rethinking the protocol design for data networking in such challenging environments.

\nosection{Cross-layer Measurement Study on Mobile Networks}
This type of work typically requires access to the low level (L1/2) information.
As studied in \cite{liu2008experiences}, TCP performance is not significantly influenced by wireless channel data rate but rather the queuing effect primarily due to the presence of large buffers in 3G networks. 
The work \cite{tso2012mobility} presents the first public report on a large-scale empirical study on the performance of commercial mobile HSPA (3.5G) networks. The key relevant finding is that the throughput performance does not monotonically decrease with increased mobility level when below 100 \kmh. 
The study \cite{merz2014performance} shown that the performance of LTE remains robust up to 200 \kmh and the SNR is the most important factor to ensure reliable operation in terms of higher order of modulation and coding schemes (MCS) and rank (number of streams).
The authors in \cite{huang2013depth} found that the high queuing delay (and its variance) in LTE networks often cause TCP congestion window to collapse upon a single packet loss, or fail to adapt fast enough and thus under-utilize the bandwidth. 
The work \cite{xu2014end} reveal burstiness pattern of packets arrival due to the polling duty cycle of the radio driver in mobile devices. 
Our work extends these findings by conducting a in-depth handover-centric study and quantifying its impact at different mobility level up to 350 \kmh on high-speed rails. \\

\nosection{Large-scale Mobile Network Usage and Performance Characterization}
The study \cite{falaki2010diversity} characterizes application usage and network traffic based on user demographics from 255 users. 
The work \cite{huang2010anatomizing} reveals that the application performance difference can be attributed to device type, operating system and web software (compression modes, concurrency). 
The authors in \cite{xu2011identifying} collected one-week's data from a tier-1 network’s UMTS core network to characterize their genre, geographic, popularity, co-occurrence, diurnal and mobility patterns. 
It was discovered \cite{chen2012network} that the mobile application performance can be enhanced by CDN-optimized initial congestion window, but may also be degraded due to the suboptimal design of application-level protocol. 
The first large real-world LTE packet trace was collected in \cite{huang2013depth} via a monitor point between eNB and EPC to analyze the flow profile (\eg size, duration, rate and concurrency) as well as the segmented network latency. 
The analysis \cite{nikravesh2014mobile} demonstrates that there is significant variance in key performance both within and across carrier at different location and time-of-day.

%% file: concl.tex
\section{Conclusion}

We perform an in-depth measurement study of HSR networking performance by examining
a wide range of factors including TCP performance metrics, flow characteristics, application breakdown, and network usage patterns. In particular, we quantitatively investigate the impact of handovers on HSR networking performance, and compare two representative TCP variants: CUBIC and BBR. Our identified performance issues are often times attributed to the upper-layer protocol design (\eg BBR's underestimation of $RTprop$), and how they interact with lower-layer characteristics (\eg TCP's unawareness of high-frequency handovers and their ``near effect''). Our insights gained from the study guides us to design a simple yet effective BBR-based congestion control solution to improve the data transfer over HSR.
In our future work, we plan to utilize our measurement findings to design transport protocol mechanisms that are more friendly to extreme mobility. We will also develop mechanisms that leverage the path diversity of multiple carriers to boost the robustness of HSR connectivity.

%% file: ms.bbl
\begin{thebibliography}{10}

\bibitem{hsr}
High-speed rail.
\newblock \url{https://en.wikipedia.org/wiki/High-speed_rail}.

\bibitem{ushsr}
High-speed rail in the united states.
\newblock
  \url{https://en.wikipedia.org/wiki/High-speed_rail_in_the_United_States}.

\bibitem{xiao2014tcp}
Qingyang Xiao, Ke~Xu, Dan Wang, Li~Li, and Yifeng Zhong.
\newblock Tcp performance over mobile networks in high-speed mobility
  scenarios.
\newblock In {\em IEEE ICNP}, 2014.

\bibitem{li2017longitudinal}
Li~Li, Ke~Xu, Dan Wang, Chunyi Peng, Kai Zheng, Rashid Mijumbi, and Qingyang
  Xiao.
\newblock A longitudinal measurement study of tcp performance and behavior in
  3g/4g networks over high speed rails.
\newblock {\em IEEE/ACM Transactions on Networking}, 2017.

\bibitem{fxh3}
China launches upgraded high-speed trains, with wi-fi.
\newblock
  \url{https://gbtimes.com/china-launches-upgraded-high-speed-trains-wi-fi}.

\bibitem{ha2008cubic}
Sangtae Ha, Injong Rhee, and Lisong Xu.
\newblock Cubic: a new tcp-friendly high-speed tcp variant.
\newblock {\em ACM SIGOPS Operating Systems Review}, 42(5), 2008.

\bibitem{bbr}
Bbr congestion control algorithm.
\newblock \url{https://github.com/google/bbr}.

\bibitem{he2016high}
Ruisi He, Bo~Ai, Gongpu Wang, Ke~Guan, Zhangdui Zhong, Andreas~F Molisch, Cesar
  Briso-Rodriguez, and Claude~P Oestges.
\newblock High-speed railway communications: From gsm-r to lte-r.
\newblock {\em IEEE Vehicular Technology Magazine}, 11(3), 2016.

\bibitem{25.913}
Universal mobile telecommunications system (umts); lte; requirements for
  evolved utra (e-utra) and evolved utran (e-utran).
\newblock \url{http://www.3gpp.org/DynaReport/25913.htm}.
\newblock 3GPP TR 25.913 version 8.0.0 Release 8 (2009-01-02).

\bibitem{russell1995interchannel}
Mark Russell and Gordon~L Stuber.
\newblock Interchannel interference analysis of ofdm in a mobile environment.
\newblock In {\em IEEE VTC}, 1995.

\bibitem{yang2012doppler}
Yaoqing Yang, Pingyi Fan, and Yongming Huang.
\newblock Doppler frequency offsets estimation and diversity reception scheme
  of high speed railway with multiple antennas on separated carriages.
\newblock In {\em IEEE WCSP}, 2012.

\bibitem{palat2009lte}
Sudeep Palat and Ph~Godin.
\newblock The lte network architecture: a comprehensive tutorial.
\newblock {\em The UMTS Long Term Evolution: From Theory to Practice. John
  Wiley \& Sons}, 2009.

\bibitem{dimou2009handover}
Konstantinos Dimou, Min Wang, Yu~Yang, Muhammmad Kazmi, Anna Larmo, Jonas
  Pettersson, Walter Muller, and Ylva Timner.
\newblock Handover within 3gpp lte: design principles and performance.
\newblock In {\em IEEE VTC Fall}, 2009.

\bibitem{36.300}
Lte; evolved universal terrestrial radio access (e-utra) and evolved universal
  terrestrial radio access network (e-utran); overall description; stage 2.
\newblock \url{http://www.3gpp.org/dynareport/36300.htm}.
\newblock 3GPP TS 36.300 version 8.12.0 Release 8 (2010-04-28).

\bibitem{li2011china}
Xing Li, Congxiao Bao, Maoke Chen, Hong Zhang, and Jianping Wu.
\newblock The china education and research network (cernet) ivi translation
  design and deployment for the ipv4/ipv6 coexistence and transition.
\newblock Technical report, 2011.

\bibitem{li2016mobileinsight}
Yuanjie Li, Chunyi Peng, Zengwen Yuan, Jiayao Li, Haotian Deng, and Tao Wang.
\newblock Mobileinsight: Extracting and analyzing cellular network information
  on smartphones.
\newblock In {\em ACM MobiCom}, 2016.

\bibitem{kwan2009proportional}
Raymond Kwan, Cyril Leung, and Jie Zhang.
\newblock Proportional fair multiuser scheduling in lte.
\newblock {\em IEEE Signal Processing Letters}, 16(6), 2009.

\bibitem{luan2013fading}
Fengyu Luan, Yan Zhang, Limin Xiao, Chunhui Zhou, and Shidong Zhou.
\newblock Fading characteristics of wireless channel on high-speed railway in
  hilly terrain scenario.
\newblock {\em International Journal of Antennas and Propagation}, 2013.

\bibitem{hsrnetdat}
\url{http://soar.group/projects/hsrnet}.

\bibitem{fxh1}
Meet china's newest high-speed train -- the fuxing hao.
\newblock
  \url{http://www.atimes.com/article/meet-chinas-newest-high-speed-train-fuxing-hao/}.

\bibitem{fxh2}
Speed limit rockets to 350 km/h.
\newblock
  \url{http://www.chinadaily.com.cn/business/2017-09/22/content_32327165.htm}.

\bibitem{qxdm}
Qxdm professional™ qualcomm extensible diagnostic monitor.
\newblock
  \url{https://www.qualcomm.com/documents/qxdm-professional-qualcomm-extensible-diagnostic-monitor}.

\bibitem{jiang2012tackling}
Haiqing Jiang, Yaogong Wang, Kyunghan Lee, and Injong Rhee.
\newblock Tackling bufferbloat in 3g/4g networks.
\newblock In {\em ACM IMC}, 2012.

\bibitem{gettys2011bufferbloat}
Jim Gettys and Kathleen Nichols.
\newblock Bufferbloat: Dark buffers in the internet.
\newblock {\em ACM Queue}, 9(11), 2011.

\bibitem{chen2013measurement}
Yung-Chih Chen, Yeon-sup Lim, Richard~J Gibbens, Erich~M Nahum, Ramin Khalili,
  and Don Towsley.
\newblock A measurement-based study of multipath tcp performance over wireless
  networks.
\newblock In {\em ACM IMC}, 2013.

\bibitem{zhang2002characteristics}
Yin Zhang, Lee Breslau, Vern Paxson, and Scott Shenker.
\newblock On the characteristics and origins of internet flow rates.
\newblock In {\em ACM SIGCOMM}, 2002.

\bibitem{chen2012network}
Xian Chen, Ruofan Jin, Kyoungwon Suh, Bing Wang, and Wei Wei.
\newblock Network performance of smart mobile handhelds in a university campus
  wifi network.
\newblock In {\em ACM IMC}, 2012.

\bibitem{huang2013depth}
Junxian Huang, Feng Qian, Yihua Guo, Yuanyuan Zhou, Qiang Xu, Z~Morley Mao,
  Subhabrata Sen, and Oliver Spatscheck.
\newblock An in-depth study of lte: effect of network protocol and application
  behavior on performance.
\newblock In {\em ACM SIGCOMM}, 2013.

\bibitem{halepovic2012can}
Emir Halepovic, Jeffrey Pang, and Oliver Spatscheck.
\newblock Can you get me now?: estimating the time-to-first-byte of http
  transactions with passive measurements.
\newblock In {\em Proceedings of the 2012 Internet Measurement Conference},
  pages 115--122. ACM, 2012.

\bibitem{john2008trends}
Wolfgang John, Sven Tafvelin, and Tomas Olovsson.
\newblock Trends and differences in connection-behavior within classes of
  internet backbone traffic.
\newblock In {\em International Conference on Passive and Active Network
  Measurement}, pages 192--201. Springer, 2008.

\bibitem{cardwell2017bbr}
Neal Cardwell, Yuchung Cheng, C~Stephen Gunn, Soheil~Hassas Yeganeh, et~al.
\newblock Bbr: congestion-based congestion control.
\newblock {\em Communications of the ACM}, 60(2), 2017.

\bibitem{jacobson1988congestion}
Van Jacobson.
\newblock Congestion avoidance and control.
\newblock In {\em ACM SIGCOMM}, 1988.

\bibitem{xu2004binary}
Lisong Xu, Khaled Harfoush, and Injong Rhee.
\newblock Binary increase congestion control (bic) for fast long-distance
  networks.
\newblock In {\em IEEE INFOCOM}, 2004.

\bibitem{brakmo1995tcp}
Lawrence~S. Brakmo and Larry~L. Peterson.
\newblock Tcp vegas: End to end congestion avoidance on a global internet.
\newblock {\em IEEE Journal on selected Areas in communications}, 13(8), 1995.

\bibitem{mascolo2001tcp}
Saverio Mascolo, Claudio Casetti, Mario Gerla, Medy~Y Sanadidi, and Ren Wang.
\newblock Tcp westwood: Bandwidth estimation for enhanced transport over
  wireless links.
\newblock In {\em ACM MobiCom}, 2001.

\bibitem{tan2006compound}
Kun Tan, Jingmin Song, Qian Zhang, and Murad Sridharan.
\newblock A compound tcp approach for high-speed and long distance networks.
\newblock In {\em IEEE INFOCOM}, 2006.

\bibitem{alrshah2014comparative}
Mohamed~A Alrshah, Mohamed Othman, Borhanuddin Ali, and Zurina~Mohd Hanapi.
\newblock Comparative study of high-speed linux tcp variants over high-bdp
  networks.
\newblock {\em Journal of Network and Computer Applications}, 43, 2014.

\bibitem{winstein2013stochastic}
Keith Winstein, Anirudh Sivaraman, Hari Balakrishnan, et~al.
\newblock Stochastic forecasts achieve high throughput and low delay over
  cellular networks.
\newblock In {\em USENIX NSDI}, 2013.

\bibitem{zaki2015adaptive}
Yasir Zaki, Thomas P{\"o}tsch, Jay Chen, Lakshminarayanan Subramanian, and
  Carmelita G{\"o}rg.
\newblock Adaptive congestion control for unpredictable cellular networks.
\newblock In {\em ACM SIGCOMM}, 2015.

\bibitem{lu2015cqic}
Feng Lu, Hao Du, Ankur Jain, Geoffrey~M Voelker, Alex~C Snoeren, and Andreas
  Terzis.
\newblock Cqic: Revisiting cross-layer congestion control for cellular
  networks.
\newblock In {\em ACM HotMobile}, 2015.

\bibitem{xie2017accelerating}
Xiufeng Xie, Xinyu Zhang, and Shilin Zhu.
\newblock Accelerating mobile web loading using cellular link information.
\newblock In {\em ACM MobiSys}, 2017.

\bibitem{leong2017tcp}
Wai~Kay Leong, Zixiao Wang, and Ben Leong.
\newblock Tcp congestion control beyond bandwidth-delay product for mobile
  cellular networks.
\newblock In {\em ACM CoNEXT}, 2017.

\bibitem{park2018exll}
Shinik Park, Jinsung Lee, Junseon Kim, Jihoon Lee, Sangtae Ha, and Kyunghan
  Lee.
\newblock Exll: an extremely low-latency congestion control for mobile cellular
  networks.
\newblock In {\em ACM CoNEXT}, 2018.

\bibitem{rula2016ips}
John~P Rula, Fabi{\'a}n~E Bustamante, and David~R Choffnes.
\newblock When ips fly: A case for redefining airline communication.
\newblock In {\em ACM HotMobile}, 2016.

\bibitem{rula2018mile}
John~P Rula, James Newman, Fabi{\'a}n~E Bustamante, Arash~Molavi Kakhki, and
  David Choffnes.
\newblock Mile high wifi: A first look at in-flight internet connectivity.
\newblock In {\em WWW}, 2018.

\bibitem{jang20093g}
Keon Jang, Mongnam Han, Soohyun Cho, Hyung-Keun Ryu, Jaehwa Lee, Yeongseok Lee,
  and Sue~B Moon.
\newblock 3g and 3.5 g wireless network performance measured from moving cars
  and high-speed trains.
\newblock In {\em ACM MICNET}, 2009.

\bibitem{li2018measurement}
Li~Li, Ke~Xu, Tong Li, Kai Zheng, Chunyi Peng, Dan Wang, Xiangxiang Wang, Meng
  Shen, and Rashid Mijumbi.
\newblock A measurement study on multi-path tcp with multiple cellular carriers
  on high speed rails.
\newblock In {\em ACM SIGCOMM}, 2018.

\bibitem{liu2008experiences}
Xin Liu, Ashwin Sridharan, Sridhar Machiraju, Mukund Seshadri, and Hui Zang.
\newblock Experiences in a 3g network: interplay between the wireless channel
  and applications.
\newblock In {\em ACM MobiCom}, 2008.

\bibitem{tso2012mobility}
Fung~Po Tso, Jin Teng, Weijia Jia, and Dong Xuan.
\newblock Mobility: A double-edged sword for hspa networks: A large-scale test
  on hong kong mobile hspa networks.
\newblock {\em IEEE Transactions on Parallel and Distributed Systems}, 23(10),
  2012.

\bibitem{merz2014performance}
Ruben Merz, Daniel Wenger, Damiano Scanferla, and Stefan Mauron.
\newblock Performance of lte in a high-velocity environment: A measurement
  study.
\newblock In {\em ACM AllThingsCellular}, 2014.

\bibitem{xu2014end}
Yin Xu, Zixiao Wang, Wai~Kay Leong, and Ben Leong.
\newblock An end-to-end measurement study of modern cellular data networks.
\newblock In {\em PAM}. Springer, 2014.

\bibitem{falaki2010diversity}
Hossein Falaki, Ratul Mahajan, Srikanth Kandula, Dimitrios Lymberopoulos,
  Ramesh Govindan, and Deborah Estrin.
\newblock Diversity in smartphone usage.
\newblock In {\em ACM MobiSys}, 2010.

\bibitem{huang2010anatomizing}
Junxian Huang, Qiang Xu, Birjodh Tiwana, Z~Morley Mao, Ming Zhang, and Paramvir
  Bahl.
\newblock Anatomizing application performance differences on smartphones.
\newblock In {\em ACM MobiSys}, 2010.

\bibitem{xu2011identifying}
Qiang Xu, Jeffrey Erman, Alexandre Gerber, Zhuoqing Mao, Jeffrey Pang, and
  Shobha Venkataraman.
\newblock Identifying diverse usage behaviors of smartphone apps.
\newblock In {\em ACM IMC}, 2011.

\bibitem{nikravesh2014mobile}
Ashkan Nikravesh, David~R Choffnes, Ethan Katz-Bassett, Zhuoqing~Morley Mao,
  and Matt Welsh.
\newblock Mobile network performance from user devices: A longitudinal,
  multidimensional analysis.
\newblock In {\em PAM}, volume~14. Springer, 2014.

\end{thebibliography}
